\documentclass[aps,nofootinbib,twocolumn,groupedaddress,floatfix,eqsecnum]{revtex4-1}

\usepackage{amsmath,amssymb,bm,comment}
\usepackage{graphicx}
\usepackage{epsfig}
\usepackage{epstopdf}
\usepackage{hyperref}

\usepackage{rotating}
\usepackage[mathscr]{eucal}

\begin{document}

\title{More Holographic Berezinskii-Kosterlitz-Thouless Transitions}
\author{Kristan Jensen}
\email{\tt kristanj@u.washington.edu}
\affiliation{Department of Physics, University of Washington, Seattle,
WA 98195-1560, USA}
\date{\today}
\begin{abstract}
We find two systems via holography that exhibit quantum Berezinskii-Kosterlitz-Thouless (BKT) phase transitions. The first is the ABJM theory with flavor and the second is a flavored $(1,1)$ little string theory.  In each case the transition occurs at nonzero density and magnetic field. The BKT transition in the little string theory is the first example of a quantum BKT transition in (3+1) dimensions. As in the ``original'' holographic BKT transition in the D3/D5 system, the exponential scaling is destroyed at any nonzero temperature and the transition becomes second order. Along the way we construct holographic renormalization for probe branes in the ABJM theory and propose a scheme for the little string theory.  Finally, we obtain the embeddings and (half of) the meson spectrum in the ABJM theory with massive flavor.
\end{abstract}

\maketitle

\section{Introduction}
\label{intro}

Holography~\cite{Maldacena:1997re,Witten:1998qj,Gubser:1998bc} has become a cornerstone in the study of strongly interacting theories.  Its use has extended far beyond its strict range of applicability, affording simple geometric pictures for the physics of confinement~\cite{Witten:1998zw}, chiral symmetry breaking~\cite{Sakai:2004cn}, and phase transitions~\cite{Mateos:2006nu}.  Indeed, holography has found many more applications to heavy-ion physics, including the study of transport in strongly coupled plasmas~\cite{Son:2007vk} and the energy loss of hard partons~\cite{Herzog:2006gh,Gubser:2006bz}.  More recently there has been a flood of work realizing condensed matter phenomena via holography.  These have included the Fermi gas at unitarity~\cite{Son:2008ye,Balasubramanian:2008dm}, the quantum Hall effect~\cite{Davis:2008nv,Fujita:2009kw,Bergman:2010gm}, superfluids~\cite{Gubser:2008px,Hartnoll:2008vx}, non-Fermi liquids~\cite{Liu:2009dm,Cubrovic:2009ye,Faulkner:2009wj,Hartnoll:2009ns}, and quantum critical points~\cite{Jensen:2010vd,Jensen:2010ga,Iqbal:2010eh,D'Hoker:2010rz}.  Reviews of many of these applications can be found in~\cite{Hartnoll:2009sz,Herzog:2009xv,McGreevy:2009xe}.

The study of critical phenomena is of central importance in the condensed matter community.  In addition to its many storied successes~\cite{Wilson:1973jj} there are some unsolved problems surrounding the role of strongly interacting quantum critical points and strange metals~\cite{sachdev-2009}.  These are relevant in candidates for the theory of high $T_c$ superconductors.  It is a driving hope that holography can successfully describe these systems.  A first step toward this goal is to identify the classes of phase transitions that are natural in holographic theories.  Unfortunately, most of these transitions are rather boring from this point of view: they are usually first-order~\cite{Hawking:1982dh,Witten:1998zw,Mateos:2006nu,Kirsch:2004km} or second-order with mean-field exponents~\cite{Karch:2007br,Gauntlett:2009dn,Gauntlett:2009bh}.  This mean-field scaling is actually expected rather than surprising; there is a large $N$ parameter that suppresses quantum fluctuations in both the field~\cite{Yaffe:1981vf} and gravitational theories.  This represents a challenge to the study of more interesting phase transitions at large $N$: in additional to identifying them~\cite{Buchel:2009mf,Franco:2009if}, we should also understand \emph{why} they exist in the first place in the sense of~\cite{Rosenstein:1988dj}.

The first holographic example of a continuous non-second-order transition embedded in string theory was found in the D3/D5 system~\cite{Jensen:2010ga}.  There is a novel chiral quantum transition in that system with the scaling of the Berezinskii-Kosterlitz-Thouless (BKT) phase transition~\cite{Berezinskii,Kosterlitz:1973xp}.  Recall that transitions of the BKT type are between disordered and \emph{quasi}-ordered phases in two dimensions.  BKT transitions are rather special: their existence is intimately related to the Coleman-Mermin-Wagner theorem~\cite{Coleman:1973ci,Mermin:1966fe,Hohenberg:1967zz}.  The two-point function of the order parameter in the disordered phase exhibits a correlation length that scales with $\exp(c/\sqrt{T-T_c})$ near the critical temperature $T_c$~\cite{Kosterlitz:1974sm}.  Similarly, the free energy of the quasi-ordered phase differs from that of the ordered by an amount that scales as $\exp (-c/\sqrt{T_c-T})$.

The transition in the D3/D5 system is novel in that it has BKT scaling in an \emph{ordered} phase.  For this reason we termed it a holographic BKT transition in~\cite{Jensen:2010ga}, since it happens in a different context from the original BKT transition.  Before going on, we note two more interesting features of the critical point in the D3/D5 system.  First, the exponential scaling of the order parameter is lost for any nonzero temperature, for which the chiral transition becomes second-order with mean-field exponents~\cite{Jensen:2010ga,Evans:2010hi}.  Second, there are relics of the BKT scaling at nonzero temperature: the critical temperature of the transition scales exponentially~\cite{Iqbal:2010eh}.

We do not often stumble across new types of phase transitions.  As a result, we should ask ourselves a number of questions: (i.) Can we identify other field theories that realize holographic BKT transitions, particularly in theories without a gravitiational dual? (ii.) What is the field theory mechanism that triggers these transitions? (iii.) Can we identify an effective theory near criticality?  In this work we attack the first question by identifying more examples.  However, our goal is not to simply enumerate a list of theories with holographic BKT transitions.  Rather, we seek to develop a classification program toward the end of answering questions (ii.) and (iii.).

\begin{figure}
\includegraphics[scale=.8]{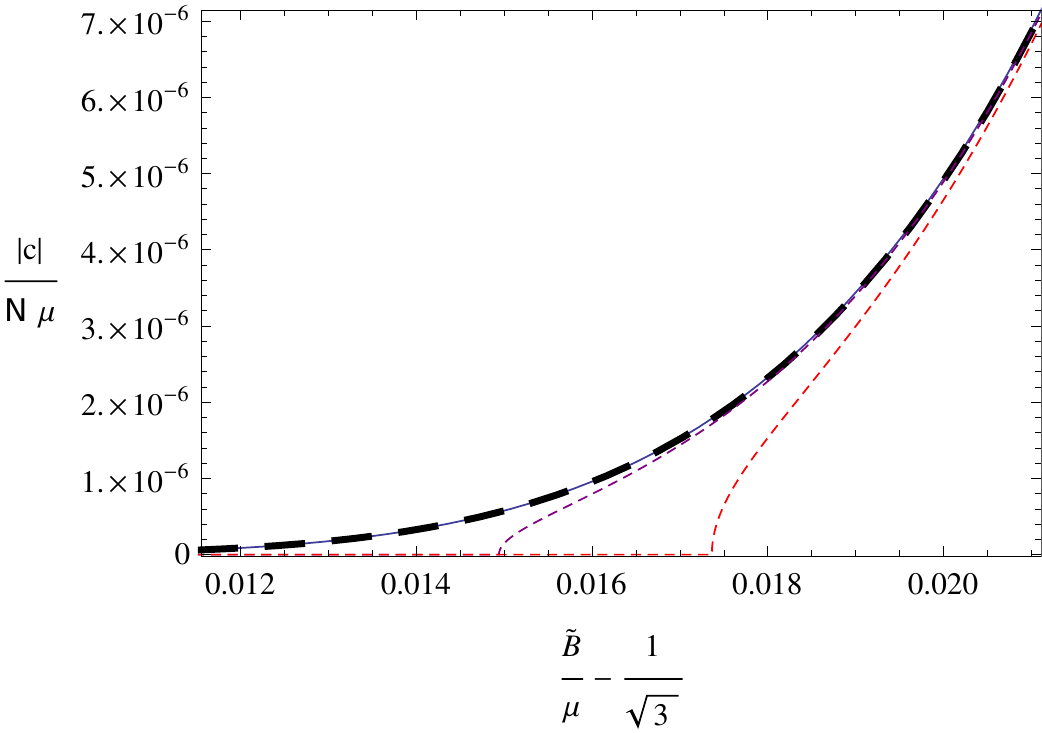}
\caption{
\label{cPlot}
A plot of the condensate in the flavored little string theory as a function of magnetic field at zero and finite temperature near the zero-temperature transition. The dashed black line indicates zero-temperature numerical data and
the solid blue line our prediction, Eq.~(\ref{cPredict}). The color dashed curves represent numerical data
at background entropy densities of $s=10^{-24}/(2\pi)^4g_s^2R$ (left) and $10^{-22}/(2\pi)^4g_s^2R$ (right). At any nonzero temperature, the condensate scales with a mean-field exponent near the transition and then asymptotes to the BKT scaling at large magnetic field.
}
\end{figure}

In this work we study flavored ABJM theory~\cite{Aharony:2008ug} and a flavored $(1,1)$ little string theory~\cite{Seiberg:1997zk} in detail.  These are supersymmetric theories that we study holographically.  In particular, we identify a holographic BKT transition in both systems at nonzero density and magnetic field.  The gravitational description of these systems is given in terms of a probe brane minimizing its worldvolume action in a fixed geometry.  In Refs.~\cite{Jensen:2010vd,Jensen:2010ga} we studied similar probes describing $(3+1)$ and $(2+1)$ dimensional field theories.  The $(3+1)$ dimensional system was also studied in~\cite{Evans:2010iy}.  There the competition between a nonzero density and magnetic field gave rise to chiral quantum transitions: the transition in the $(3+1)$ dimensional theory was second-order with mean-field scaling, while the same system in $(2+1)$ dimensions exhibits exponential scaling.  In this work, the flavored ABJM theory also lives in $(2+1)$ dimensions but the flavored little string theory is the first example of a theory in $(3+1)$ dimensions with BKT scaling.  For dramatic effect and summary, we have included a plot of the order parameter in the broken phase of that theory in Fig.~\ref{cPlot}.

The gravitational mechanism for the holographic BKT transition is clear.  It occurs due to the violation of the Breitenlohner-Freedman (BF) bound~\cite{Breitenlohner:1982jf} in the infrared region by the scalar field dual to an order parameter.  On general grounds presented in~\cite{Kaplan:2009kr} this violation was expected to produce BKT scaling.  However, there are only a handful of known settings where the BF bound is violated in a controlled setting.  The first was the D3/D5 system, the second in extremal asymptotically AdS$_4$ dyonic black holes~\cite{Iqbal:2010eh}, and the third in this work\footnote{While we were finishing this work, S. Pal released a paper~\cite{Pal:2010gj} that also studies holographic BKT transitions in the D$p$/D$q$ systems.  Their results at nonzero density and magnetic field agree with ours.}.  While this mechanism seems similar to that of a second-order Landau phase transition, the bulk field actually represents an infinite tower of field theory modes.  At the transition, an infinite number of these modes destabilize.  This situation is quite unnatural within the Landau theory.  To us these results suggest that we need a new paradigm of phase transitions to describe the infrared physics of these systems.

While we have not yet found such a paradigm, we have identified a picture for some sufficient conditions to trigger a holographic BKT transition.  We do this by considering a large class of theories without holographic BKT transitions.  In particular, we add flavor to the field theories dual to the near-horizon $p$-brane geometries.  These are interacting theories in $(p+1)$ dimensions with running couplings (except for $p=3$, corresponding to $\mathcal{N}=4$ super Yang-Mills (SYM) theory) but with a \emph{generalized conformal symmetry}.  Operators typically carry different dimensions than their canonical ones, a feat accomplished by redefining them with powers of the couplings.  After turning on a density and magnetic field, we find that the only theories that realize holographic BKT transitions are the D3/D5 system and the flavored little string theory of this work.  Combined with flavored ABJM, these are also the only theories for which the \emph{generalized} dimension of the density and magnetic field equal each other.  We do not think that this is a coincidence.  From these results we conjecture that a holographic BKT transition can be triggered in a generalized conformal theory after adding equal-dimension control parameters with some other constraints.  See our discussion in Sec.~\ref{discuss} for more details.

There are a few other cute results that we establish while studying these BKT transitions.  The first is in the ABJM theory.  In order to properly study chiral symmetry breaking in that theory, we find it convenient to obtain the brane embeddings dual to the theory with supersymmetric massive flavor.  While we do not explicitly verify kappa-symmetry, we perform a few consistency checks that these branes are indeed supersymmetric.  After holographically renormalizing the bulk theory, we find that the free energy and condensates of the field theory vanish.  For another check, we obtain half of the meson masses of the dual theory.  These are suppressed from the quark mass by a factor of the 't Hooft coupling of the ABJM theory, $m_{\rm meson}\sim m_{\rm quark}/\sqrt{\lambda}.$  We also find hints of a broken $SO(5)$ invariance in the meson masses, much like in the D3/D7 system~\cite{Kruczenski:2003be}.  Finally, in the course of regularizing the little string theory we also holographically renormalize the D3/D3 system at nonzero density.  The most interesting result here is a new Weyl anomaly of the dual theory proportional to $\mu^2$, where $\mu$ is the chemical potential for baryon density.

The outline of this work follows.  In Sec.~\ref{setups} we describe the probe brane setups we use, beginning with flavored ABJM theory in Sec.~\ref{abjmSetup}.  First we review the ABJM theory and the probe branes dual to the addition of massless flavor.  We go on to construct the embeddings dual to supersymmetric massive flavor and compute some the meson masses of that theory.  Next, we review the $(1,1)$ little string theory and the addition of flavor in Sec.~\ref{ldSetup}.  In Sec.~\ref{holoRG} we holographically renormalize probe branes in the ABJM theory and propose a scheme for a holographic map in the little string theory.  The bulk of our results are in Sec.~\ref{results}, where we analytically and numerically identify the holographic BKT transitions in these theories.  We also perform numerics for the little string theory, measuring the fate of exponential scaling at nonzero temperature in Sec.~\ref{finiteT}.  Finally, we discuss our results in the context of the D$p$/D$q$ systems and generalized conformal symmetry in Sec.~\ref{discuss}.

\section{The Brane Setups}
\label{setups}
\subsection{ABJM with flavor}
\label{abjmSetup}
The geometry supported by a large stack of $N$ M2 branes in M-theory probing the orbifold singularity of $\mathbb{C}^4/\mathbb{Z}_k$  has a near-horizon limit of AdS$_4\times \mathbb{S}^7/\mathbb{Z}_k$,
\begin{eqnarray}
\nonumber
g&&= \frac{R^2}{4}g_{\rm AdS_4}+R^2 g_{\mathbb{S}^7/\mathbb{Z}_k}\\
&&(R/l_{\rm Pl})^6=32\pi^2 k N,
\end{eqnarray}
with $kN$ units of 4-form flux through the AdS$_4$ factor.  The field theory living on the branes is $\mathcal{N}=6$ $U(N)\times U(N)$ superconformal Chern-Simons theory at level $k$~\cite{Aharony:2008ug}.  This is the celebrated ABJM theory.  At levels $k=1,2$, the theory has enhanced $\mathcal{N}=8$ supersymmetry~\cite{Aharony:2008ug,Benna:2008zy}.  Moreover, the theory at $N=2$ is identical to the Bagger-Lambert theory~\cite{Bagger:2006sk,Bagger:2007jr} proposed a few years ago as the worldvolume theory on a stack of M2 branes.

The orbifolded circle inside $\mathbb{S}^7/\mathbb{Z}_k$ has a size $R/k$ which is small when $N\ll k^5$.  In this limit the M2 branes are well-described by the compactification of AdS$_4\times \mathbb{S}^7/\mathbb{Z}_k$ on this circle, i.e. by type IIA string theory on AdS$_4\times \mathbb{CP}^3$.  In string frame, this background is
\begin{eqnarray}
\nonumber
\label{abjmBackground}
g&=&\frac{R^3}{4k}(g_{\rm AdS_4} + 4g_{\mathbb{CP}^3}), \\
e^{2\phi}&=&\frac{R^3}{k^3}, \,\,\, F_4 = \frac{3}{8}R^3 \text{vol(AdS}_4\text{)}, \,\,\, F_2 =k J,
\end{eqnarray}
where we have set $\alpha'=1$.  The two-form $F_2$ appears in the 10d description because the reduction is on a non-trivially fibered circle inside $\mathbb{S}^7/\mathbb{Z}_k$; the connection associated with the fibration is the potential for the K\"ahler form on $\mathbb{CP}^3$, $J$.  This background was first identified in~\cite{Nilsson:1984bj} as the dimensional reduction of the AdS$_4\times \mathbb{S}^7$ solution of 11d supergravity on the diagonal circle in $\mathbb{S}^7$.

We therefore have a IIA description of the ABJM theory at large $k$ and $N$ when $k^5\gg N$ and when the 10d radius of curvature is large, i.e. $R^3/4k\propto \sqrt{N/k}\gg 1$.  Most of the applications of this duality have centered on the conjectured integrability of the ABJM theory by studying the dimensions of large spin operators~\cite{Arutyunov:2008if,Stefanski:2008ik,Grignani:2008is,Minahan:2008hf,Gaiotto:2008cg,Gromov:2008qe,Bak:2008cp,Gromov:2009tv}.  Additionally, there has been some work done adding flavors to the ABJM theory~\cite{Hikida:2009tp,Gaiotto:2009tk,Hohenegger:2009as,Ammon:2009wc}.  This addition is realized in the bulk by adding a small number of probe D-branes~\cite{Karch:2002sh} that wrap some cycles inside the $\mathbb{CP}^3$.  

It is a short exercise to show that adding a single fundamental hypermultiplet to one of the gauge groups breaks the R-symmetry of the theory to $SO(3)$~\cite{Hikida:2009tp,Gaiotto:2009tk,Hohenegger:2009as} and so the supersymmetry to $\mathcal{N}=3$.  There is an additional $SO(3)_{\chi}$ global symmetry preserved when the flavor is massless so that the global symmetry is $SO(3)_{\rm R}\times SO(3)_{\chi}\times U(1)_{\rm B}$, where the last $U(1)$ symmetry is a baryon number symmetry.  By supersymmetry the dual brane setup should have a single D6 brane wrapping a 3-cycle inside of $\mathbb{CP}^3$.  Additionally, this cycle should have a $\mathbb{Z}_2$ fundamental group, so that there can be two different Wilson lines on the brane corresponding to adding the flavor to one gauge group or the other~\cite{Douglas:1996sw,Hikida:2009tp}.

Up to $SU(4)$ rotations of $\mathbb{CP}^3$ there is a single $SO(4)$ invariant 3-cycle, $\mathbb{RP}^3$.  This cycle also has a $\mathbb{Z}_2$ fundamental group, making it a consistent candidate for the correct flavor brane embedding.  Indeed, a D6 brane wrapping AdS$_4\times \mathbb{RP}^3$ preserves twelve supercharges of supersymmetry~\cite{Hikida:2009tp}.  In the probe limit massless flavored ABJM theory is a superconformal $\mathcal{N}=3$ theory, so this is the correct dual.
 
However, our ultimate goal is to study $SO(3)_{\chi}$ chiral-symmetry breaking in this theory.  The embeddings dual to the chirally broken phase of this theory will deviate from the supersymmetric embedding.  The deformation should be in the same direction as that of the embeddings dual to \emph{massive} flavor, which have not yet been obtained.  Thus, in order to correctly study $SO(3)_{\chi}$ breaking we elect to study the addition of massive flavor.  This is actually not that hard.  

In the bulk we will have a dual flavor brane with a non-trivial embedding and zero field strength for the $U(1)$ gauge field on the brane.  The non-trivial part of this problem is that these embeddings will deform the $\mathbb{RP}^3$ as a function of the radial coordinate.   We will therefore look more carefully at the embedding of $\mathbb{RP}^3$ inside of $\mathbb{CP}^3$.  Following~\cite{Hikida:2009tp}, we find it instructive to construct $\mathbb{CP}^3$ inside of $\mathbb{C}^4$.  We will use this construction to describe $\mathbb{RP}^3$ and its deformations.

\subsubsection{Massive flavor}

We begin with $N$ M2 branes on $\mathbb{C}^4$, which we parametrize with complex coordinates $z_i,i=1-4$ as
\begin{eqnarray}
\nonumber 
z_1&=&r \cos\xi \cos\frac{\theta_1}{2}e^{i\frac{\chi_1+\phi_1}{2}}, \,\,\, z_2=r\cos\xi\sin\frac{\theta_1}{2}e^{i\frac{\chi_1-\phi_1}{2}}, \\
z_3&=&r \sin \xi \cos \frac{\theta_2}{2}e^{i\frac{\chi_2+\phi_2}{2}}, \,\,\, z_4=r\sin\xi\sin\frac{\theta_2}{2}e^{i\frac{\chi_2-\phi_2}{2}},
\end{eqnarray}
where $r$ is the radial coordinate of $\mathbb{C}^4$ and the angular variables have the domains $\xi\in [0,\frac{\pi}{2}], \theta_i\in [0,\pi], \chi_i\in[0,4\pi), \phi_i\in [0,2\pi)$.  Defining new angular coordinates $y$ and $\psi$ by 
\begin{equation}
\chi_1=2y+\psi, \,\,\, \chi_2=2y-\psi,
\end{equation}
we see that the diagonal circle of $\mathbb{C}^4$  is given by $y\in[0,2\pi)$ along which the $\mathbb{Z}_k$ orbifold acts.  The orbifold simply identifies $y\sim y+\frac{2\pi}{k}$, so that we write the metric of the angular part $\mathbb{S}^7/\mathbb{Z}_k$ as
\begin{equation}
g_{\mathbb{S}^7/\mathbb{Z}_k}=\frac{1}{k^2}(dy+A)^2+g_{\mathbb{CP}^3},
\end{equation}
which simply says that $\mathbb{S}^7/\mathbb{Z}_k$ can be written as a Hopf fibration over $\mathbb{CP}^3$ with connection $A$.  In these coordinates we have
\begin{equation}
2A=\cos 2\xi d\psi+\cos^2\xi\cos\theta_1 d\phi_1+\sin^2\xi\cos\theta_2 d\phi_2,
\end{equation}
and
\begin{widetext}
\begin{equation}
g_{\mathbb{CP}^3}=d\xi^2+\cos^2\xi\sin^2\xi\left( d\psi+\frac{\cos\theta_1}{2}d\phi_1-\frac{\cos\theta_2}{2}d\phi_2 \right)^2+\frac{1}{4}\cos^2\xi (d\theta_1^2+\cos^2\theta_1d\phi_1^2)+\frac{1}{4}\sin^2\xi (d\theta_2^2+\cos^2\theta_2d\phi_2^2).
\end{equation}
\end{widetext}
The connection $A$ is the potential for the K\"ahler form on $\mathbb{CP}^3$, $dA=J$.

The $\mathbb{RP}^3$ inside of $\mathbb{CP}^3$ we consider is invariant under the simultaneous $SU(2)$ rotations of $(z_1,z_2)$ and $(z_3,z_4)$ (the bulk realization of the $SO(3)_{\rm R}$ symmetry) and is given by
\begin{equation}
\mathcal{C}_3: \,\,\, \theta_1=\theta_2=\theta, \,\,\,\, \phi_1=-\phi_2=\phi, \,\,\,\, \xi=\frac{\pi}{4}.
\end{equation}
The induced metric on this space is
\begin{equation}
g_{\mathcal{C_3}}=\frac{1}{4}\left( (d\psi+\cos\theta d\phi)^2+d\theta^2+\sin^2\theta d\phi^2 \right),
\end{equation}
where the coordinates have domains $\theta\in [0,\pi], \phi\in [0,2\pi), \psi\in [0,2\pi)$.  This is indeed the metric for a round unit $\mathbb{RP}^3$.  It is also invariant under a second $SU(2)$ that simultaneously rotates $(z_1,z_4)$ and $(z_3,z_2)$.  This $SU(2)_2\approx SO(3)_{\chi}$ is the bulk realization of the extra chiral symmetry of the massless theory.

Of course this is not the only 3-cycle invariant under the $SO(3)_{\rm R}$.  There is an obvious $SO(3)_{\rm R}$ invariant 3-cycle
\begin{equation}
\mathcal{C}_3:\,\,\, \theta_1=\theta_2=\theta,\,\,\,\,\phi_1=-\phi_2=\phi,\,\,\,\,\xi=\xi_0,
\end{equation}
which becomes $\mathbb{RP}^3$ for $\xi_0=\pi/4$.  The $\xi$-direction is transverse to this cycle as are two other directions related by $SO(3)_{\chi}$ transformations.  The fluctuations of the cycle under these three directions transform as a triplet under $SO(3)_{\chi}$.  The dual operator is the $SO(3)_{\chi}$ triplet hypermultiplet mass operator $\bar{\psi}\psi+\text{SUSY}$~\cite{Hikida:2009tp}.  We find it convenient to consider brane embeddings where the brane slips off of $\mathbb{RP}^3$ in the $\xi$-direction.

There is a useful change-of-coordinates for this problem.  Define $y,\rho$ by
\begin{equation}
\xi=\frac{\pi}{4}+\frac{1}{2}\text{arctan}\,\frac{y}{\rho}, \,\,\, r^2=y^2+\rho^2.
\end{equation}
In these coordinates the Poinc\'are horizon of AdS$_4$ is located at $y=\rho=0$ and the boundary of AdS$_4$ at either $y,\rho\rightarrow\infty$.  Then a D6 brane that wraps $\mathcal{C}_3$ with $\xi=\xi(r)$ can be described with an embedding $y=y(\rho)$ and an induced metric
\begin{equation}
\text{P}[g]=\frac{R^3}{4k}\left(r^2 g_{2,1}+\frac{d\rho^2(1+y'^2)+4\rho^2 g_{\mathcal{C}}}{r^2} \right),
\end{equation}
where the internal space has a metric
\begin{equation}
g_{\mathcal{C}}=\frac{1}{4}\left( \left(1+\frac{y^2}{\rho^2}\right)(d\theta^2+\sin^2\theta d\phi^2)+(d\psi+\cos\theta d\phi)^2 \right).
\end{equation}
We recognize this space as a squashed unit $\mathbb{RP}^3$ with a squashing parameter $y/\rho$,
\begin{equation}
g_{\mathcal{C}}=g_{\mathbb{RP}^3}+\frac{1}{4} \frac{y^2}{\rho^2}(d\theta^2+\sin^2\theta d\phi^2).
\end{equation}
When $y=0$ the internal space is a round $\mathbb{RP}^3$ and as $y/\rho\rightarrow\infty$ the brane ends smoothly as $\det \text{P}[g]\rightarrow 0$.  This change-of-coordinates is similar to the one in~\cite{Kruczenski:2003be} that has made many computations in probe brane systems tractable.  Notably, this squashing only preserves the $U(1)$ in $SO(3)_{\chi}$ that rotates $\psi$.  The global symmetry of the dual theory is then $SO(3)_R\times U(1)_{\chi}\times U(1)_{\rm B}$.

Before obtaining the embeddings dual to massive flavor, we need one more ingredient.  There is a non-trivial Chern-Simons term on the D6 branes that must be included in order to obtain the correct equations of motion for the embedding, the integral of $C_7$ on the worldvolume.  The pullback of the $C_7$ to the brane is~\cite{Hikida:2009tp}
\begin{equation}
\label{C7}
\text{P}[C_7]=\frac{R^9}{2^8k^2}r^2\sin\theta dx^0\wedge dx^1\wedge dx^2\wedge dr\wedge d\psi\wedge d\theta\wedge d\phi,
\end{equation}
and the brane action is
\begin{equation}
S_6=-T_6\int d^7\xi e^{-\phi}\sqrt{-\text{P}[g]}+T_6\int \text{P}[C_7]
\end{equation}
in the absence of any worldvolume field strength.  This Lagrangian for the embedding function $y$ is rather complicated.  Integrating over the internal space and dividing by the volume of the field theory directions $x^{\mu}$, it is proportional to
\begin{equation}
\mathcal{L}_6\propto -\rho\sqrt{\rho^2+y^2}\sqrt{1+y'^2}+\frac{y^2(\rho+yy')}{2\sqrt{\rho^2+y^2}}.
\end{equation}
The equation of motion for this action has a remarkably simple solution, $y=$constant.  We claim that these embeddings describe the ABJM theory with massive supersymmetric flavor: we have consistent supergravity solutions dual to the field theory where we have turned on a source for the hypermultiplet mass operator.  Indeed, as we show in Sec.~\ref{abjmRG}, (i.) there is a renormalization scheme under which the dual extremum has zero free energy and (ii.) the expectation value of the field dual to $y$ vanishes.  Both of these results are important consistency checks with the supersymmetry of the dual theory.  Yet another check, the meson spectrum, is mostly computed in Appendix~\ref{appSpectrum} and is reviewed below.

\subsubsection{The meson spectrum}
\label{abjmMesons}

We obtain the meson spectrum of flavored ABJM theory by studying the fluctuations of worldvolume fields around the embedding $y=m$.  In order to be a small fluctuation these fluctuations must be small and well-behaved everywhere on the brane, particularly near $\rho=0$.  This is a boundary condition on the fluctuations.  Fourier-transforming in the field theory directions $x^{\mu}$, the fluctuations will only obey a normalization condition near the AdS$_4$ boundary for particular momenta $-k^2$.  These are the masses(-squared) of the mesons of the dual field theory.

For our choice of parametrization there are three classes of worldvolume fields, (i.) the transvese scalar $y$ and those related by $SO(3)_{\chi}$ symmetry, (ii.) the $U(1)$ gauge field $A$, and (iii.) a neutral fermion $\Psi$.  As we show in Appendix~\ref{appSpectrum}, the fluctuations of the gauge field can be broken up further into three types, a vector and two different types of scalars.  We elect to study the bosonic mesons on the presumption that supersymmetry relates their masses to the fermionic mesons.

The supercharges of the theory fill a triplet of the $SO(3)_{\rm R}$ symmetry.  These commute with the $SO(3)_{\chi}$ chiral symmetry, which is broken to $U(1)_{\chi}$ by turning on a mass.  All states in a given supermultiplet then have the same $U(1)_{\chi}$ charge.  Our mesonic multiplets can then be obtained by acting with the supercharges on a scalar state with spin $\frac{j}{2}$ under the $SO(3)_{\rm R}$ that is annihilated by the $\bar{Q}$'s.  Before going on, we will use the notation $\left(\frac{j}{2};\frac{j'}{2},\frac{m'}{2}\right)$ to denote the quantum numbers of a meson, where $j,j'$ are $SO(3)_{\rm R}$ and $SO(3)_{\chi}$ spins respectively and $\frac{m'}{2}$ is the $U(1)_{\chi}$ charge.  Also, $j,j,',$ and $m'$ must be even, which is related in the bulk to the fact that the internal space $\mathcal{C}_3$ is a $\mathbb{Z}_2$ identification of a squashed 3-sphere.

We are not able to obtain the full meson spectrum of the theory.  We are, however, able to obtain several sectors.  We describe their computation in Appendix~\ref{appSpectrum} and simply summarize here.  First, we solve the fluctuation spectrum of the transverse scalar $y$ and find the meson masses
\begin{equation}
\label{yMesons2}
-k^2=\frac{m^2}{4}(2+j+|m'|+4n)(4+j+|m'|+4n),
\end{equation}
in the $\left(\frac{j}{2};\frac{j}{2},\frac{m'}{2}\right)$ represenation for any positive integer $n$.  The operators dual to these fields have dimension $2+\frac{j}{2}$.  The $j=0$ operator is the hypermultiplet mass operator $\bar{\psi}\psi+ \text{SUSY}$, which has dimension $2$ as expected.  The other transverse scalars are related to $y$ by $SO(3)_{\chi}$ symmetry but they have $U(1)_{\chi}$ charges $\pm 1$.  The masses of these dual mesons are then given by Eq.~(\ref{yMesons2}), but with $m'$ shifted by $\mp 2$.  The net effect is that for $|m'|\leq j$ there are three sets of degenerate meson spectra with masses Eq.~(\ref{yMesons2}) and that there are extra mesons with $U(1)_{\chi}$ charge $\pm\left(\frac{j}{2}+1\right)$ and masses
\begin{equation}
-k^2=m^2(1+j+2n)(2+j+2n).
\end{equation}
We are also able to obtain the masses of the vector mesons, dual to eigenmodes of a type of gauge field.  These obey the same equation as the transverse scalar and so have the same spectrum as Eq.~(\ref{yMesons2}).  These also live in the $\left(\frac{j}{2};\frac{j}{2},\frac{m'}{2}\right)$ representation and are related to operators of dimensions $\Delta = 2+\frac{j}{2}$.

There are a few sectors that we were not able to obtain.  The first were another set of scalars in the $\left(\frac{j}{2};\frac{j}{2},\frac{m'}{2}\right)$ representation dual to another fluctuation of the gauge field.  We did, however, study this sector numerically and found good agreement for several $j,m'$ with the masses Eq.~(\ref{yMesons2}).  We can also obtain the zero momentum bulk wavefunctions, from which we measure the dimensions of the dual operators, which are $2+\frac{j}{2}$.  

The real mess is found in the last set of scalars, dual to fluctuations of the gauge field in the internal directions.  The difficulty is that there are three different vector harmonics on the internal space $\mathcal{C}_3$ that generally mix under the remaining symmetries of the problem.  There are a few exceptions, including the $j=0$ multiplet, for which the harmonic analysis is easy.  These fields will feature prominently for the rest of the paper, as it is turned on when we have a background density and magnetic field.  We therefore write their functional form for reference
\begin{eqnarray}
\nonumber
\label{j=0}
m'=0:&&\, A=d\psi+\cos\theta d\phi,  \\
m'=\pm 2:&&\, A=e^{\pm i\psi}(\pm id\theta+ \sin\theta d\phi).
\end{eqnarray}
These fields are dual to the $SO(3)_{\chi}$ triplet scalar mass operator $(\bar{Q}Q-\bar{\tilde{Q}}\tilde{Q},\tilde{Q}Q,\bar{\tilde{Q}}\bar{Q})$.  This operator has dimension $1$ in the supersymmetric theory, so some care needs to be taken in the bulk to ensure that we impose supersymmetric boundary conditions: the normalizable term of the near-boundary expansion behaves as a source and the non-normalizable is related to the expectation values of the dual operator.  Unfortunately, we were not able to find the spectrum of these, nor the other internal gauge fields.

The meson masses we have obtained, while only representing half of the full spectrum, exhibit a huge degeneracy: all states of the same $j+|m'|+4n$ have the same mass.  This result is reminiscent of a broken $SO(5)$ degeneracy in the D3/D7 system~\cite{Kruczenski:2003be}, where all the meson states were furnished by representations of $SO(5)$ but not all states in an irreducible representation were degenerate.  We are finding hints of a similar effect here: the induced metric on the D6 branes is conformal to $\mathbb{E}^{2,1}\times \mathcal{C}_4$,
\begin{equation}
\text{P}[g]=(\rho^2+m^2)g_{2,1}+\frac{1}{\rho^2+m^2}(\underbrace{d\rho^2+4\rho^2 g_{\mathcal{C}_3}}_{=g_{\mathcal{C}_4}}),
\end{equation}
where the $\mathcal{C}_4$ is a squashed, orbifolded 4-dimensional space.  The conformal factor then depends on a coordinate on $\mathcal{C}_4$.  Near the boundary the squashing vanishes and $\mathcal{C}_4$ also has local $SO(5)$ isometry.  We speculate that the hints of broken $SO(5)$ invariance are arising from the global isometries of $\mathcal{C}_4$ near the boundary.  The combination of the conformal factor and the squashing presumably breaks them to $SO(3)_{\rm R}\times U(1)_{\chi}$.  It would be interesting to obtain the remaining meson spectra and more precisely understand what is happening on $\mathcal{C}_4$.

\subsubsection{Magnetic field and density}

We can now study chiral symmetry breaking in this theory.  We will probe the phase diagram of this theory with a baryon density and magnetic field.  The baryon current is dual to the $U(1)$ gauge field living on the brane; we therefore add a baryon density and magnetic field to the field theory by turning on a field strength on the brane
\begin{equation}
F=\frac{R^3}{4k}\left(A_0'(\rho)\,d\rho\wedge dx^0+B\, dx^1\wedge dx^2\right),
\end{equation}
where we have normalized the field strength in the same units as the background metric Eq.~{\ref{abjmBackground}}.  The field $A_0$ is dual to the density operator and $B$ is the magnetic field on the brane and in the field theory.  We will consider brane embeddings that are translationally invariant in the field theory directions (012) while only turning on the field strength above and the transverse scalar $y=y(\rho)$.  The equivalent ansatz\"e in the field theory is that the correct extrema in the chirally broken phase are translationally invariant and singlets under $SO(3)_R$ symmetry.

These ansatz\", however, are not consistent.  There is a Chern-Simons term on the brane
\begin{equation}
S_{\rm CS}=\frac{T_6}{6}\int \text{P}[C_1]\wedge F\wedge F\wedge F,
\end{equation}
where $C_1$ is the potential for 2-form flux $F_2$.  The pullback of $C_1$ to our branes is
\begin{equation}
\text{P}[C_1]=-k\frac{y}{2\sqrt{\rho^2+y^2}}(d\psi+\cos\theta d\phi)
\end{equation}
so that this Chern-Simons term is linear in a flux on the internal space $\mathcal{C}_3$.  Careful study of this term shows that a density and magnetic field induces the $(0;1,0)$ mode of the internal gauge field when the embedding is non-trivial.  A consistent Ansatz therefore has a field strength
\begin{equation}
\frac{4kF}{R^3}=A_0'(\rho)\,d\rho\wedge dx^0+B\,dx^1\wedge dx^2+d(A_{\psi}(r)(d\psi+\cos\theta d\phi)).
\end{equation}
Recall that this internal gauge field is dual to the scalar mass operator.  This field and $y$ transform in the same representation of the global symmetry group and so the dual operators are both good order parameters for chiral symmetry breaking.  It is interesting that both of them are turned on in this problem.

Since our embeddings will have internal gauge fields turned on the brane action
\begin{equation}
S_6=-T_6\int d^{7}\xi e^{-\phi}\sqrt{-\text{P}[g]+F}+T_6\sum_i\int \text{P}[C_i]\wedge e^{F}
\end{equation}
 will have three non-trivial Chern-Simons terms, namely those with $C_1$, $C_3$, and $C_7$.  For our ansatz\"e the brane action becomes
\begin{widetext}
\begin{equation}
\label{abjmS}
S_6=-\underbrace{T_6\frac{R^9}{2^4k^2}\text{vol}[\mathbb{RP}^3]}_{\equiv \mathcal{N}}\text{vol}[\mathbb{E}^{2,1}]\int d\rho \left[\rho r\sqrt{1+y'^2-A_0'^2+\frac{r^4}{\rho^2}A_{\psi}'^2}\sqrt{1+\frac{B^2}{r^4}}\sqrt{1+A_{\psi}^2}-\frac{y^2(\rho+yy')}{2r}-\frac{yA_0'BA_{\psi}}{r}+r^3A_{\psi}A_{\psi}'   \right],
\end{equation}
\end{widetext}
where we have used
\begin{equation}
\text{P}[C_3]=\frac{R^3}{8}r^3 dx^{0}\wedge dx^1\wedge dx^2,
\end{equation}
and have defined the positive orientation $(x^0,x^1,x^2,\rho,\psi,\theta,\phi)$.  For convenience, from here on we will consider the action \emph{density} $S_6/\text{vol}[\mathbb{E}^{2,1}]$ and give it the same name $S_6$.  Now we note that the field $A_0$ only appears in the action Eq.~(\ref{abjmS}) through radial derivatives.  We therefore eliminate it in favor of a constant of the motion,
\begin{equation}
\label{abjmDens}
\frac{\partial\mathcal{L}}{\partial A_0'(\rho)}=\mathcal{N}A_0'\left[\frac{\rho r\sqrt{1+\frac{B^2}{r^4}}\sqrt{1+A_{\psi}^2}}{\sqrt{1+y'^2-A_0'^2+\frac{r^4}{\rho^2}A_{\psi}'^2}}+\frac{yBA_{\psi}}{r} \right]=d.
\end{equation}
which we show in Sec.~\ref{abjmRG} is the baryon density of the dual theory.  In order to find solutions at fixed density, we algebraically solve Eq.~(\ref{abjmDens}) for $A_0'$ and then Legendre-transform the action Eq.~(\ref{abjmS}) to fixed density,
\begin{widetext}
\begin{equation}
\label{abjmS2}
\tilde{S}_6=-\mathcal{N}\int d\rho \left[ \frac{1}{r}\sqrt{1+y'^2+\frac{r^4}{\rho^2}A_{\psi}'^2}\sqrt{\rho^2(1+A_{\psi}^2)(B^2+r^4)+(\tilde{d}r-BA_{\psi}y)^2}-\frac{y^2(\rho+yy')}{2r}+r^3A_{\psi}A_{\psi}'  \right],
\end{equation}
\end{widetext}
where $\tilde{d}=d/\mathcal{N}$ is a rescaled density.  Solutions that extremize this (fairly nasty) action will correspond to extrema in the canonical ensemble of the dual theory.

\subsection{Little string theory}
\label{ldSetup}
The geometry supported a large stack of $N$ D5-branes in type IIB string theory has a near-horizon limit~\cite{Itzhaki:1998dd}
\begin{eqnarray}
\nonumber
g&=&\frac{r}{R}g_{5,1}+\frac{R}{r}(dr^2+r^2g_{\mathbb{S}^3}), \\
e^{\phi}&=& g_s\frac{r}{R}, \,\,\,  R=\sqrt{g_sN},
\end{eqnarray}
where $g_s$ is the string coupling at infinity before taking the near-horizon limit and we have set $\alpha'=1$.  There are also $N$ units of 3-form field strength on the 3-sphere sourced by the 5-branes.  Type IIB string theory on this background is dual to the theory on the 5-branes, which in the $g_s\rightarrow 0$ limit is the $(1,1)$ little string theory on a stack of $N$ NS5-branes~\cite{Itzhaki:1998dd}.  This theory is not a local quantum field theory and rather has a number of stringy features, including string-like excitations, T-duality, and a Hagedorn spectrum at high temperatures~\cite{Seiberg:1997zk}.  For these reasons and because it lives in six rather than ten dimensions, it is called a ``little string theory.''  For obvious reasons, the dual geometry is termed a ``linear dilaton'' background.  The little string theory is strongly coupled in the ultraviolet and infrared free; in fact, the infrared theory is the maximally supersymmetric (5+1)-dimensional super-Yang-Mills theory on the D5 branes.  The full theory can also be obtained via deconstruction~\cite{ArkaniHamed:2001ie}.

The linear dilaton background is conformal to $\mathbb{E}^{6,1}\times \mathbb{S}^3$, i.e. to flat space.  As a result, there is some difficulty in building a holographic map from bulk quantities to field theory observables\footnote{There has been some work renormalizing asymptotically linear dilaton backgrounds from type II theories in~\cite{Marolf:2007ys,Cotrone:2007qa}.}.  The 5-brane background is rather special in this respect: the near-horizon region of the $p$-brane backgrounds are generically conformal to AdS$_{p+2}\times \mathbb{S}^{8-p}$, which yields a dictionary through holographic renormalization~\cite{Kanitscheider:2009as,Benincasa:2009ze}.

Nevertheless we can learn a great deal about the little string theory from bulk physics.  Bulk thermodynamics correspond to field theory thermodynamics; two-point functions in the field theory can be obtained by scattering wave packets off of the center of the geometry~\cite{Barbon:2007za,Bertoldi:2009yi}.  In this work, we revisit an item on this list: we can add flavor to the field theory by embedding probe branes in the dual geometry~\cite{Karch:2002sh,Canoura:2005fc}.  

There are a number of supersymmetric ways to add flavor to the little string theory, all of which are realized with dual probe branes that wrap cycles with four Neumann-Dirichlet directions with respect to the 5-branes.  These brane setups and thus the dual field theories preserve eight supercharges with flavor living on various codimension defects in the field theory.  The example we will look at in this work is that of codimension-2 flavor.  The weak-coupling, small $N$ brane embedding has $N$ D5 branes along the 012345 directions and flavor D5 branes along the 012367 directions.  In the large $N$ limit the flavor branes wrap a $(5+1)$ dimensional cycle inside the linear dilaton background including a circle inside the transverse 3-sphere. We write the 10d metric as
\begin{equation}
\label{ldZeroTMetric}
g=\frac{\sqrt{\rho^2+y^2}}{R}g_{5,1}+\frac{R}{\sqrt{\rho^2+y^2}}(d\rho^2+dy^2+\rho^2d\theta^2+y^2d\phi^2),
\end{equation}
where the radial coordinate $r$ is related to these coordinates by $r=\sqrt{\rho^2+y^2}$, the supersymmetric brane embeddings are specified by $x^4=x^5=\phi =\rm constant$ and $y=m$, where $m$ is proportional to the mass of the dual flavor.  As in the ABJM theory, the 5-branes at the bottom of the geometry are located at $\rho=y=0$.  The boundary of the geometry is found as $\rho$ or $y\rightarrow \infty$.  The field theory has a $U(1)\times U(1)$ R-symmetry, realized in the bulk as the $U(1)\times U(1)$ isometries of the two circles $\theta$ and $\phi$.  Both symmetries are chiral.

The field theory has an additional $U(1)$ baryon number symmetry which rotates the fundamental scalars and fermions.  The baryon symmetry current is dual to a $U(1)$ gauge field living on the flavor branes.  Denoting its field strength as $F$, the action of the $N_f$ flavor branes is given by
\begin{equation}
S_5=-N_f T_5 \int d^6\xi e^{-\phi} \sqrt{-\text{P}[g]+F}+ S_{\rm CS},
\end{equation}
where the branes have a tension $T_5$, the $\xi$ are coordinates on the worldvolume, $\text{P}[g]$ is the induced metric on the branes, and $S_{\rm CS}$ represents the couplings of these branes to the Ramond-Ramond form fields.

As before, we will probe the phase diagram of this theory at nonzero baryon density and magnetic field.  We therefore turn on a field strength
\begin{equation}
F=A_0'(\rho)d\rho\wedge dt+B\, dx^1\wedge dx^2.
\end{equation}
We will consider brane embeddings that preserve transitional invariance in the field theory directions (0123), are flat in the defect directions (45), and simply wrap the $\theta$-circle with $\phi=$constant.  The only freedom left in these ans\"atze is to let the transverse position $y$ of the branes depend on the radial coordinate $\rho$ as $y=y(\rho)$.  The induced metric on the branes is
\begin{equation}
\text{P}[g]=\frac{\sqrt{\rho^2+y^2}}{R}g_{3,1}+\frac{R}{\sqrt{\rho^2+y^2}}(d\rho^2(1+y'^2)+\rho^2d\theta^2).
\end{equation}
The Chern-Simons terms with these ans\"atze are cubic in worldvolume fields that we do not turn on, and so we can consistently ignore them.  Rescaling the spatial and radial coordinates by factors of $R$ the brane action becomes
\begin{eqnarray}
\nonumber
\label{ldDBI}
S_5=-&&\underbrace{N_f T_5R^6\text{vol}[\mathbb{S}^1]}_{\equiv \mathcal{N}}\text{vol}[\mathbb{E}^{3,1}] \\
&&\hspace{.4cm}\times\int d\rho \, \rho\sqrt{1+y'^2-A_0'^2}\sqrt{1+\frac{B^2}{r^2}}.
\end{eqnarray}
We pause to note that, at zero magnetic field, this is the same action for the D3/D3 system at nonzero density.  Much of what we do (excepting the magnetic field) can then be recast in terms of that system.  Also, as in our discussion in ABJM we will hereafter write of the action \emph{density} $S_5/\text{vol}[\mathbb{E}^{3,1}]$ and give it the same name $S_5$.

Since the field $A_0$ only appears in the action Eq.~(\ref{ldDBI}) through derivatives, we eliminate it in favor of a constant of the motion.  Ordinarily this constant is interpreted as the baryon density~\cite{Kobayashi:2006sb,Karch:2007br}. In this case, however, the gauge field has a near-boundary falloff $A_0=\mu \log r+O(1)$, where $\mu$ is naturally interpreted as the chemical potential in the field theory.  For this system (as with the D3/D3 system), the constant of the motion is instead proportional to $\mu$,
\begin{equation}
\label{ldConst}
\frac{\partial \mathcal{L}}{\partial A_0'(\rho)}=\frac{\mathcal{N}\rho A_0'\sqrt{1+\frac{B^2}{r^2}}}{\sqrt{1+y'^2-A_0'^2}}=\mathcal{N}\mu.
\end{equation}
Solving Eq.~(\ref{ldConst}) for $A_0'$ and Legendre-transforming the bulk action to fixed $\mu$, we find the new action
\begin{equation}
\label{ldAction}
\tilde{S}_5=-\mathcal{N}\int d\rho \sqrt{1+y'^2}\sqrt{\mu^2+\rho^2\left( 1+\frac{B^2}{r^2} \right)},
\end{equation}
Curiously the bulk action at fixed $\mu$ Eq.~(\ref{ldAction}) corresponds to the field theory action at fixed density.  We will return to this in Sec.~\ref{ldProbe}.

There is one more control parameter that we can trivially turn on: the analogue of the scalar mass we looked at in the ABJM theory Eq.~(\ref{j=0}).  In the D3/D3 system as well as here, the dual field is the zero-mode of the internal gauge field on the $\theta$-circle.  Turning it on with a field strength
\begin{equation}
F=A_0'(\rho)d\rho\wedge dt +B\, dx^1\wedge dx^2 + A_{\theta}'(\rho)d\rho\wedge d\theta,
\end{equation}
the brane action becomes
\begin{equation}
\label{ldDBI2}
S_5=-\mathcal{N}\int d\rho \,\rho\sqrt{1+y'^2-A_0'^2+\frac{r^2}{\rho^2}A_{\theta}'^2}\sqrt{1+\frac{B^2}{r^2}}.
\end{equation}
As before, we note that $A_0$ and $A_{\theta}$ both appear through derivatives so that we have two constants of the motion.  The reader can verify that $A_{\theta}$ has the large-$\rho$ falloff $A_{\theta}=M \log \rho+O(1)$, so that the constants are
\begin{eqnarray}
\nonumber
 \frac{\partial\mathcal{L}}{\partial A_0'(\rho)}&=&\frac{\mathcal{N}\rho A_0'\sqrt{1+\frac{B^2}{r^2}}}{\sqrt{1+y'^2-A_0'^2+\frac{r^2}{\rho^2}A_{\theta}'^2}}=\mathcal{N}\mu, \\
- \frac{\partial\mathcal{L}}{\partial A_{\theta}'(\rho)}&=&\frac{\mathcal{N}r^2 A_{\theta}'\sqrt{1+\frac{B^2}{r^2}}}{\rho\sqrt{1+y'^2-A_0'^2+\frac{r^2}{\rho^2}A_{\theta}'^2}}=\mathcal{N}M.
\end{eqnarray}
Solving for $A_0'$ and $A_{\theta}'$ in terms of $\mu$ and $M$ we find
\begin{eqnarray}
\nonumber
A_0'&=&\frac{\mu\sqrt{1+y'^2}}{\sqrt{\mu^2+\rho^2\left( 1+\frac{B^2-M^2}{r^2} \right)}}, \\
\frac{A_{\theta}'}{R^2}&=&\frac{M \rho^2\sqrt{1+y'^2}}{r^2\sqrt{\mu^2+\rho^2\left( 1+\frac{B^2-M^2}{r^2} \right)}},
\end{eqnarray}
Legendre transforming the bulk action Eq.~(\ref{ldDBI2}) with respect to both $A_0'$ and $A_{\theta}'$ we find the new action
\begin{equation}
\label{ldAction2}
\tilde{S}_5=-\mathcal{N}\int d\rho\sqrt{1+y'^2}\sqrt{\mu^2+\rho^2\left( 1+\frac{B^2-M^2}{r^2} \right)}.
\end{equation}
Two comments are in order: (i.) the effect of adding a scalar mass can be completely solved in terms of a ``new'' magnetic field with magnitude $\tilde{B}=\sqrt{B^2-M^2}$, (ii.) reality of the action informs us that there is an upper bound on the scalar mass, $M^2_{\rm max}=\mu^2 + B^2$, that can be described with probe branes.

\subsection{Holographic renormalization}
\label{holoRG}
\subsubsection{Probe branes in ABJM}
\label{abjmRG}

We continue by holographically renormalizing flavored ABJM theory.  We have several tasks in this subsection: (i.) to show that the embeddings we propose are dual to massive flavor have zero free energy, (ii.) to obtain one-point functions at nonzero density and magnetic field, and (iii.) determine the proper operator normalization for the transverse scalars.  We will achieve all of these tasks via holographic renormalization.

The action Eq.~(\ref{abjmS2}) diverges near the AdS$_4$ boundary.  This infrared divergence is related to an ultraviolet divergence of the dual theory.  We consistently renormalize it by diffeomorphism-invariantly regulating the bulk theory.  We do this by first solving the equations of motion for $y$ and $A_{\psi}$ near the boundary,
\begin{equation}
y=m+\sum_{n=1}^{\infty} \frac{y_n}{\rho^n}, \,\,\, A_{\psi}=\sum_{n=1}^{\infty}\frac{A_n}{\rho^n},
\end{equation}
where the higher $y_n$ are recursively determined by $m$ and $y_1$ and the higher $A_n$ are determined by $A_1$ and $A_2$.  Next we integrate the action on an arbitrary solution up to a very large cutoff $\rho=\Lambda$,
\begin{equation}
\tilde{S}_{6,\Lambda}=-\mathcal{N}\int^{\Lambda}d\rho \,\rho^2+\text{finite}.
\end{equation}
The brane action then has a single divergence from the volume of AdS$_4$.  We regulate it by adding a local counterterm on the cutoff slice
\begin{equation}
\tilde{S}_{\rm CT}=\mathcal{N}\frac{\sqrt{-\gamma}}{3}|_{\rho=\Lambda},
\end{equation}
where $\gamma$ is the induced metric on the cutoff slice.  For an action with a more general divergence pattern we would add a number of local, (cutoff-)diffeomorphism-invariant counterterms on the cutoff slice.  

We define our renormalized action by adding the counterterm to the action Eq.~(\ref{abjmS2}) and then taking the cutoff to infinity,
\begin{equation}
\tilde{S}_{6,\rm ren}=\lim_{\Lambda\rightarrow\infty}\left[ \tilde{S}_{5,\Lambda}+S_{\rm CT}\right].
\end{equation}
This bulk action is finite and convergent on-shell; from it we can obtain correlators of the dual theory in the usual way.  We begin with our first goal, the free energy of the embeddings $y=m$.  The renormalized action for these branes is just
\begin{equation}
\tilde{S}_{6,\rm ren}=-\mathcal{N}\frac{m^3}{6},
\end{equation}
which is nonzero.  However, we have missed one subtlety: we have not specified the correct infrared boundary condition for $\text{C}_7$.  To see this, recall that there is a term in the brane action $\int \text{P}[C_7]$ that changes under gauge transformations $C_7\mapsto C_7+dC_6$ with the boundary term $\tilde{S}\mapsto \tilde{S}+\int_{\partial}C_6$, where $\partial$ is the boundary of the worldvolume.  Gauge transformations with support at the worldvolume boundary change the boundary conditions on the bulk theory.  In particular, when $C_6$ does not vanish on the AdS$_4$ boundary it changes the source for a magnetic current in the dual theory~\cite{Witten:2003ya}.  On the other hand, a non-trivial $C_6$ at the bottom of the brane changes the infrared value of the gauge field.  We will impose the boundary condition that the gauge field vanishes there.  However, the pullback of $C_7$ in Eq.~(\ref{C7}) does not vanish at the bottom of the brane,
\begin{equation}
\text{P}[C_7]|_{\rho=0}=\frac{R^9}{2^8k^2}m^2\sin\theta dx^0\wedge dx^1\wedge dx^2\wedge dr\wedge d\psi\wedge d\theta\wedge d\phi.
\end{equation} 
In order to be consistent with our boundary conditions we therefore need to perform a gauge transformation so that $\text{P}[C_7]$ vanishes at the bottom of the brane without changing its near-boundary behavior.  This can be done with a $C_6$ that vanishes at large $r$ and at smaller $r$ becomes
\begin{equation}
C_6\sim\frac{R^9}{2^8k^2}\frac{r^3}{3}\sin\theta dx^0\wedge dx^1\wedge dx^2\wedge d\psi\wedge d\theta\wedge d\phi.
\end{equation}
This transformation adds a contribution to the bulk action proportional to $m^3$.  In fact, the renormalized action in this gauge simply vanishes.  The free energy $-\tilde{S}$ of the dual background thus vanishes as advertized.

Next we will compute one-point functions.  In order to obtain the density, let us consider the brane action before Legendre transforming Eq.~({\ref{abjmS}}).  This action diverges in the same way as its Legendre transform and so we define $S_{6,\rm ren}$ in the same way as before.  This action is equal to (minus) the free energy of the dual state in the grand canonical ensemble.  Under a small variation $A_0\mapsto A_0+\delta A_0$, $y\mapsto y+\delta y$, $A_{\psi}\mapsto A_{\psi}+\delta A_{\psi}$, the on-shell action changes by a boundary term on the cutoff slice
\begin{equation}
S_{\rm bdy}=\delta A_0\frac{\partial \mathcal{L}}{\partial A_0'}+\delta y\frac{\partial \mathcal{L}}{\partial y'}+\delta A_{\psi}\frac{\partial \mathcal{L}}{\partial A_{\psi}'}.
\end{equation}
Recalling the near-boundary expansion of $A_0$
\begin{equation}
A_0=\mu-\frac{d}{\mathcal{N}\rho}+...,
\end{equation}
and treating the leading term $A_1$ in the large-$\rho$ solution of $A_{\psi}$ as the source for the dual operator, we therefore find the one-point functions
\begin{eqnarray}
c&=&\left\langle \mathcal{O}_y\right\rangle = -\frac{\partial S_{\rm bdy}}{\partial m}=-\mathcal{N}y_1, \\
d&=&\left\langle j^0\right\rangle = \frac{\partial S_{\rm bdy}}{\partial \mu}, \\
C&=&\left\langle \mathcal{O}_{\psi}\right\rangle = -\frac{\partial S_{\rm bdy}}{\partial A_1}=-\mathcal{N}A_2.
\end{eqnarray}
However as we noted in Sec.~\ref{abjmMesons}, these boundary conditions on $A_{\psi}$ do not correspond to the supersymmetric boundary conditions where $A_2$ is the source for the dual operator.  To implement this boundary condition we need to add a total derivative to the bulk action
\begin{equation}
S_{6,\rm ren}\mapsto S_{6,\rm ren}-\frac{\mathcal{N}}{2}\int d\rho\, (\rho^3 A_{\psi}^2)'
\end{equation}
which accordingly shifts the boundary action.  The one-point function of $\mathcal{O}_{\psi}$ with these boundary conditions is then
\begin{equation}
C=\left\langle\mathcal{O}_{\psi}\right\rangle = -\frac{\partial S_{\rm bdy}}{\partial A_2}=\mathcal{N}A_1.
\end{equation}
Notably, all one-point functions vanish for the embeddings $y=m$ as advertized earlier.

Finally, we will obtain the relative normalization between the hypermultiplet mass operator $\mathcal{M}$ and $\mathcal{O}_y$.  We could follow~\cite{Karch:2005ms} and compute the two-point function $\left\langle \mathcal{O}_y(x)\mathcal{O}_{y}(0)\right\rangle$ from holographic renormalization as well as $\left\langle \mathcal{M}(x)\mathcal{M}(0)\right\rangle$ from the field theory.  We elect to take a simpler approach and compute the mass of a straight string that hangs from the bottom of the flavor brane to the D2 branes at the bottom of AdS$_2$,
\begin{equation}
M=\frac{1}{2\pi}\int_0^m d\rho \sqrt{-\text{P}[g]}=\frac{m}{\sqrt{2}}\sqrt{\frac{N}{k}}.
\end{equation}
This is the mass of the lightest charged quasiparticle in the dual theory and therefore corresponds exactly to the hypermultiplet mass.  We therefore have $\mathcal{M}=\sqrt{\frac{2k}{N}}\mathcal{O}_y$.  This relative normalization contains an important piece of physics: the quark mass ($M$) is larger than than meson masses (set by the scale $m$) by a factor of $\sqrt{N/k}$, the square root of the 't Hooft coupling of the ABJM theory.  This is precisely analogous to the mesons in the D3/D7 system (among others), where the meson masses go like $m_{\rm quark}/\sqrt{\lambda}$, where $\lambda$ is the 't Hooft coupling of the gauge theory~\cite{Kruczenski:2003be}.

\subsubsection{Probe branes in the linear dilaton background}
\label{ldProbe}

Two comments are in order before we continue.  First, because the linear dilaton background is conformal to flat space rather than to AdS, the regularization scheme we propose should be considered to be just that: a regularization scheme.  We have not rigorously renormalized the bulk theory.  As a result, our main reason for studying this regularization is to nail down the proper holographic renormalization in the D3/D3 system at nonzero density~\cite{Karch:2005ms}.  The technical details of that problem are identical to those we face in creating our regularization scheme.  

While our regularization and so our free energy for the dual theory has issues, our prescription for the chiral symmetry breaking condensate is more reasonable.  Our result for this condensate is identical to its computation in the D3/D3 system and simply returns the normalizable term in the large-$\rho$ expansion of the field $y$.  This matches the usual AdS/CFT picture where the normalizable term of a field is proportional to the vacuum expectation value of the operator sourced by its non-normalizable term.

An embedding that minimizes the action Eq.~(\ref{ldAction2}) will have the large-$\rho$ falloff $y=m\, \log \rho+y_0 + O(\log^2 \rho/\rho^2)$.  Integrating the action Eq.~(\ref{ldAction2}) up to a large radial cutoff $\Lambda$, we find
\begin{equation}
\tilde{S}_{5,\Lambda}=-\mathcal{N}\int^{\Lambda} d\rho \,\rho\left( 1+\frac{m^2+\mu^2+B^2-M^2}{2\rho^2} \right)+\text{finite}.
\end{equation}
The brane action then has a volume divergence from the first term and four logarithmic divergences from the four worldvolume fields we have turned on.  We can regulate these divergences with local counterterms on the cutoff slice $\rho=\Lambda$ in a similar manner to holographic renormalization,
\begin{equation}
\tilde{S}_{\rm CT}=\mathcal{N}\sum_{i}\tilde{\mathcal{L}}_{i}|_{\rho=\Lambda},
\end{equation}
with the counterterms 
\begin{eqnarray}
\nonumber
\tilde{\mathcal{L}}_1&=&\frac{1}{2}\sqrt{-\gamma}, \\
\nonumber
\tilde{\mathcal{L}}_2&=&\frac{1}{2\log \rho}\sqrt{-\gamma}\left( \frac{y}{\rho} \right)^2, \\
\nonumber
\tilde{\mathcal{L}}_3&=&-\frac{1}{2\log \rho}\sqrt{-\gamma}\frac{Re^{-\phi}}{g_s}A_{\mu}A^{\mu}, \\
\nonumber
\tilde{\mathcal{L}}_4&=& \frac{\log \rho}{4}\sqrt{-\gamma}F_{\mu\nu}F^{\mu\nu}, \\
\tilde{\mathcal{L}}_5&=&-\frac{1}{2\log\rho}\sqrt{-\gamma}\frac{Re^{-\phi}}{g_s}A_{\theta}^2.
\end{eqnarray}
where $\gamma$ is the induced metric on the cutoff slice and all induces are slice indices contracted with $\gamma$.  Note that in the D3/D3 system the proper counterterm for the gauge field $\tilde{\mathcal{L}}_2$ does not have a factor of the dilaton.  We then define the regulated action
\begin{equation}
\tilde{S}_{5,\rm ren}\equiv \lim_{\Lambda\rightarrow\infty}\left[ \tilde{S}_{5,\Lambda}+\tilde{S}_{\rm CT}\right].
\end{equation}

We should contrast this result with the divergences of the original action Eq.~(\ref{ldDBI2}),
\begin{equation}
S_{5,\Lambda}=-\mathcal{N}\int^{\Lambda}d\rho\, \rho\left( 1+\frac{m^2-\mu^2+B^2+M^2}{2\rho^2} \right),
\end{equation}
which differs from the divergence structure of the Legendre transformed action Eq.~(\ref{ldAction2}) in the sign of the $\mu^2$ and $M^2$ logarithmic divergences.  The Legendre transform changes the action's asymptotics and so we need slightly different slice counterterms,
\begin{equation}
S_{\rm CT}=\mathcal{N}\sum_i \mathcal{L}_i |_{\rho=\Lambda},
\end{equation}
with
\begin{eqnarray}
\nonumber \mathcal{L}_1&=&\tilde{\mathcal{L}}_1, \,\,\, \mathcal{L}_2 = \tilde{\mathcal{L}_2}, \,\,\, \mathcal{L}_4=\tilde{\mathcal{L}}_4, \\
\mathcal{L}_3&=&-\tilde{\mathcal{L}_3}, \,\,\, \mathcal{L}_5=-\tilde{\mathcal{L}}_5,
\end{eqnarray}
from which we define the regularized action
\begin{equation}
\label{ldRen}
S_{5,\rm ren}=\lim_{\Lambda\rightarrow\infty}\left[ S_{5,\Lambda}+S_{\rm CT}\right].
\end{equation}

What physics is contained in these divergences?  First, the logarithmic divergences of the bulk action correspond to a Weyl anomaly in the dual theory proportional to terms like $m^2$, $\mu^2$, \&c.  We can see this by performing an infinitesimal Weyl variation in the boundary theory, which corresponds to an infinitesimal scale transformation in the bulk.  The logarithmic terms in the bulk action transform inhomogeneously, giving rise to a change in the action under the variation~\cite{Henningson:1998gx}.  Second, some terms in the anomaly differ by a sign depending on the field theory ensemble.  

In order to make this statement more precise we need to relate bulk fields to expectation values in order to see what field theory ensembles correspond to our regularized bulk actions.  Usually the original action Eq.~(\ref{ldDBI2}) corresponds to the grand canonical ensemble and the Legendre-transformed action to the canonical.  In order to see if this is the case, we need a prescription for computing expectation values.

Even though we do not performed strict holographic renormalization for the bulk theory, let us proceed and \emph{define} one point functions in the field theory by variation of our regularized actions.  Moreover we will assume that the original action Eq.~(\ref{ldDBI2}) corresponds to the grand canonical ensemble and then show that this assumption is thermodynamically consistent.  Varying the original action Eq.~(\ref{ldRen}) with respect to the fields $y,A_0,$ and $A_{\theta}$ we find a variation of the boundary action on the cutoff slice
\begin{eqnarray}
\nonumber
\delta S_{\rm bdy}&=&\mathcal{N} [\rho\left(-\delta y\, y'+\delta A_0 A_0' - \delta A_{\theta} A_{\theta}' \right) \\
&&+ \frac{1}{\log \Lambda}(\delta y\, y-\delta A_0 A_0+\delta A_{\theta}A_{\theta})].
\end{eqnarray}
We define our one-point functions by variation of this boundary action with respect to the sources $m,\mu$, and $M$ in the $\Lambda\rightarrow \infty $ limit, for example
\begin{equation}
c=\left\langle \mathcal{O}_y\right\rangle \equiv -\frac{\delta S_{\rm bdy}}{\delta m}=-\lim_{\Lambda\rightarrow\infty}\log \Lambda \frac{\delta S_{\rm bdy}}{\delta y}.
\end{equation}
Recalling the large-$\rho$ expansion of our worldvolume fields,
\begin{eqnarray}
\nonumber
y&=&m\log \rho+y_0+..., \,\,\, A_0=\mu \log \rho+a_0+...,\\
A_{\theta}&=&M\log \rho+c_0+...
\end{eqnarray}
we therefore find our one-point functions
\begin{eqnarray}
\nonumber
\label{ldVeVs}
c&=&\left\langle \mathcal{O}_y\right\rangle\equiv -\frac{\delta S_{\rm bdy}}{\delta m} =- \mathcal{N}y_0, \\
\nonumber
d&=&\left\langle j^0\right\rangle\equiv \frac{\delta S_{\rm bdy}}{\delta \mu}=-\mathcal{N}a_0, \\
C&=&\left\langle \mathcal{O}_{\theta}\right\rangle\equiv \frac{\delta S_{\rm bdy}}{\delta M}=\mathcal{N}c_0.
\end{eqnarray}

What happens when we Legendre transform?  The transformed action Eq.~(\ref{ldAction2}) is related to the original Eq.~(\ref{ldDBI2}) by
\begin{equation}
\tilde{S}_5=S_5-\int d\rho \left(A_0'\frac{\partial \mathcal{L}}{\partial A_0'}+A_{\theta'}\frac{\partial \mathcal{L}}{\partial A_{\theta}'}  \right).
\end{equation}
Since each of the transforms gives a logarithmic divergence at large $\rho$ the counterterms required to regularize $\tilde{S}_5$ are different than for $S_5$.  Noting that $a_0$ and $c_0$ can be obtained by integrating the gauge fields,
\begin{eqnarray}
\nonumber
\label{ldDensity}
a_0&=&\lim_{\Lambda\rightarrow\infty}\left[ \int^{\Lambda}d\rho \,A_0'-\mu \log \Lambda\right], \\
c_0&=&\lim_{\Lambda\rightarrow\infty}\left[\int^{\Lambda}d\rho \,A_{\theta}'-M\log\Lambda\right],
\end{eqnarray}
the reader can show that $\tilde{S}_{5,\rm ren}$ is related to $S_{5,\rm ren}$ by
\begin{equation}
\tilde{S}_{5,\rm ren}=S_{5,\rm ren}-\mu d-M C.
\end{equation}
Since the bulk action corresponds to (minus) the free energy of the dual equilibrium state, we have
\begin{equation}
\tilde{F}=F+\mu d+M C,
\end{equation}
so that the original bulk action $S_5$ indeed corresponds to the grand canonical ensemble and $\tilde{S}_5$ to the canonical.

One nice check for these results is to compute the density in the grand canonical ensemble for the symmetric phase $c=0$.  We can do this in two different ways: (i.) by our prescription for the density in Eq.~(\ref{ldVeVs}) and (ii.) by computing the free energy $F=-S_{6,\rm ren}$ for this phase.  Thermodynamic consistency of our regularized free energy requires that $d=-\partial F/\partial \mu$.  The regularized action and dual density for the symmetric embedding $y=0$ can be evaluated by means of Eqs.~(\ref{ldRen},)(\ref{ldVeVs}), and (\ref{ldDensity}) to be
\begin{eqnarray}
\nonumber
S_{6,\rm ren}=&&\frac{r_h\sqrt{\mu^2+\tilde{B}^2+r_h^2}}{2}\\
\nonumber
&&-(\mu^2+\tilde{B}^2)\left[1-2\log\frac{r_h+\sqrt{\mu^2+\tilde{B}^2+r_h^2}}{2} \right], \\
d=&&\mu\log \frac{r_h+\sqrt{\mu^2+\tilde{B}^2+r_h^2}}{2}.
\end{eqnarray}
The reader can verify that $d=\partial S_{6,\rm ren}/\partial \mu$, so that our prescription for the free energy is indeed thermodynamically consistent in the symmetric phase, at least with respect to density.

\section{Results}
\label{results}
\subsection{ABJM with flavor}
\label{abjmResults}
Our goal in this subsection is to identify the existence and location of the BKT transition in this system.  As a result, we will not solve the full problem of obtaining embeddings, condensates, \&c in the broken phase.  The full problem, while soluble, is numerically daunting for little payoff: the final results will look very similar to those of the D3/D5 system, but the broken phase will be described by two condensates rather than one.

As in the D3/D5 system, the onset of the chiral BKT transition can be found by analyzing the stability of fluctuations around the symmetric embedding $y=A_{\psi}=0$.  These fluctuations are governed by the quadratic part of Eq.~(\ref{abjmS2}), which is equivalent to\footnote{By equivalent, we mean that there are terms of the form $f(\rho)yy'$ in the quadratic action whose contribution to the equation of motion can be compensated for by a term $-\frac{f'(\rho)}{2}y^2$.}
\begin{equation}
\label{abjmLinear}
\mathcal{L}_6\sim -\frac{\mathcal{N}}{2}\left[Y'\cdot K\cdot Y'+Y\cdot V\cdot Y\right],
\end{equation}
where
\begin{eqnarray}
Y&=&\left(\begin{array}{c}y/\rho \\ A_{\psi}\end{array}\right),\,\,\, K=\sqrt{-G}\rho^2I_2, \\
V&=&\left(\begin{array}{cc}-\rho^2-\frac{\rho^4+B^2}{\sqrt{-G}} & -\frac{\tilde{d}B}{\sqrt{-G}} \\  -\frac{\tilde{d}B}{\sqrt{-G}}&   -3\rho^2+\frac{\rho^4+B^2}{\sqrt{-G}} \end{array}\right),
\end{eqnarray}
and
\begin{equation}
\sqrt{-G}=\sqrt{\tilde{d}^2+B^2+\rho^4}
\end{equation}
is the determinant of the open string metric on the symmetric embedding.  This system behaves rather differently in two different regimes.  At large $\rho\gg \sqrt{B}$, the fields $y/\rho$ and $A_{\psi}$ decouple from each other and fluctuate as $m^2=-2$ scalars in AdS$_2$.  Fluctuations here correspond to the ultraviolet of the dual theory, while fluctuations at small $\rho\ll \sqrt{B}$ correspond to the deep infrared.  Here $y/\rho$ and $A_{\psi}$ fluctuate as coupled scalars in AdS$_2$.  The infrared eigenmodes are
\begin{equation}
\label{abjmMix}
\Phi_{\pm}=A_{\psi}+\frac{B\pm\sqrt{\tilde{d}^2+B^2} }{\tilde{d}}\frac{y}{\rho}
\end{equation}
with masses $m^2_{\pm}=\mp B/\sqrt{\tilde{d}^2+B^2}$ in the effective AdS$_2$ region.  This is the crux of the argument: while the D6 brane is always embedded in an AdS$_4$ background, the fields living on the brane see different geometries at different $\rho$.  The open string metric~\cite{Seiberg:1999vs} on the brane is dominated in the infrared by the control parameters and there becomes AdS$_2$-like (the full result is a Lifshitz-like space, but this does not affect our results).

For large density $\tilde{d}^2/B^2\geq 15$, the mode $\Phi_{\text{sgn}(B)}$ fluctuates stably in the IR.  However for $\tilde{d}^2/B^2<15$, the mass of $\Phi_{\text{sgn}(B)}$ passes below the Breitenlohner-Freedman bound for stability~\cite{Breitenlohner:1982jf} in the effective AdS$_2$, $m^2_{\rm BF}=-1/4$.  There will be an infinite, exponential hierarchy of tachyons in the symmetric embedding signalling an instability toward a chiral symmetry broken $y,A_{\psi}\neq 0$ embedding.  There is therefore a chiral quantum phase transition at
\begin{equation}
\frac{d}{|B|}_c=\mathcal{N}\sqrt{15}=\frac{N\sqrt{\frac{N}{k}}}{2^{5/2}\pi}\sqrt{15}.
\end{equation}
The appearance of the factor $N\sqrt{N/k}$ is analogous to the factor of $N\sqrt{\lambda}$ that appears in the chiral transition in the D3/D7 system~\cite{Jensen:2010vd}, where $\lambda$ is the 't Hooft coupling of that gauge theory.

We have verified the existence and location of the chiral quantum transition.  On the grounds of~\cite{Jensen:2010ga,Kaplan:2009kr}, we expect that it is of the holographic BKT type.  We can back up this claim with more rigor by sketching a derivation of the stable embeddings in the broken phase.  These embeddings determine the thermodynamics of the broken phase.  Near the transition we find the embeddings by matching solutions in two regions: a small-$\rho$ non-linear core and a large-$\rho$ linear bulk.  We begin with the core region $\rho\ll \sqrt{B}$ where the brane action is well-approximated by
\begin{eqnarray}
\nonumber
\label{abjmCore}
\tilde{S}_6=-\mathcal{N}\int &&d\rho\frac{1}{r}\sqrt{1+y'^2+\frac{r^4}{\rho^2}A_{\psi}'^2}\\
&&\times\sqrt{\rho^2B^2(1+A_{\psi}^2)+(\tilde{d}r-BA_{\psi}y)^2}.
\end{eqnarray}
Our brane embeddings extremize this action and obey the infrared boundary conditions $y(0)=0$ with $A_{\psi}$ falling off so that the action is integrable near $\rho=0$.

Importantly, this region enjoys a scaling symmetry
\begin{equation}
\rho\mapsto \xi\rho,\,\,\, y\mapsto \xi y,\,\,\, A_{\psi}\mapsto A_{\psi}.
\end{equation}
As a result we can find a family of embeddings that obey our boundary condition: for $y=f_{y}(\rho),A_{\psi}=f_{\psi}(\rho)$ a particular solution that obeys our boundary conditions, we define the solutions $y_{\xi}(\rho)=\xi f_y(\rho/\xi), A_{\psi,\xi}=f_{\psi}(\rho/\xi)$ for all $\xi\in\mathbb{R}_+$.  However, this scaling symmetry is not enough to obtain the two-parameter family of core solutions.  We need an extra boundary condition, which can only be imposed by matching to the near-AdS$_4$-boundary asymptotics.  Moving on, consider large $\rho$ in this region (near the boundary of the IR AdS$_2$), $y$ and $A_{\psi}$ mix in the same way as we noted in Eq.~(\ref{abjmMix}); there are two scalars with the asymptotics of $m^2=\pm B/\sqrt{\tilde{d}^2+B^2}$ scalars in AdS$_2$.  We are interested in embeddings close to the chiral transition, so we consider $|B|$ close to the critical value.  Defining
\begin{equation}
\alpha_{\pm}=2\sqrt{m_{\rm BF}^2-m_{\pm,\rm IR}^2}=\sqrt{-1\pm\frac{4B}{\sqrt{B^2+\tilde{d}^2}}},
\end{equation}
this region corresponds to $\alpha_{\text{sgn}(B)}$ both real and small.  Now letting $B>0$, for a particular solution $y,A_{\psi}$, the eigenmodes $\Phi_{\pm}$ have the near-AdS$_2$-boundary asymptotics
\begin{eqnarray}
\nonumber
\Phi_{+}(\rho)&\sim&\rho^{-\frac{1}{2}}\left(-a_{0}\cos\left(\frac{\alpha}{2}\log\rho\right) +\frac{2a_1}{\alpha}\sin\left(\frac{\alpha}{2}\log\rho\right)\right), \\
\Phi_-(\rho)&\sim&\rho^{-\frac{1}{2}}\left( b_0\rho^{\frac{a}{2}}+b_1\rho^{-\frac{a}{2}} \right), \,\,\,\,\, \rho\gg 1
\end{eqnarray}
where the $a_i$ and $b_i$ depend on $\alpha\equiv \alpha_+$ and we have defined $a\equiv i\alpha$.  There will be a one-parameter family of $a_i$'s and $b_i$'s that, together with scaling, generate the asymptotics of all core solutions.

To order $\alpha$, a general embedding that obeys our boundary conditions will have the large-$\rho$ asymptotics
\begin{eqnarray}
\nonumber
\label{abjmCorePhi}
\Phi_{\xi,+}(\rho)&=&\sqrt{\frac{\xi}{\rho}}\frac{2a_1}{\alpha}\sin\left(\frac{\alpha}{2}\log\frac{\rho}{\rho_1}\right), \,\,\, \rho\gg \rho_1=\xi e^{a_0/a_1}, \\
\Phi_{\xi,-}(\rho)&=&\sqrt{\frac{\xi}{\rho}}\left(b_0\xi^{-\frac{a}{2}}\rho^{\frac{a}{2}}+b_1\xi^{\frac{a}{2}}\rho^{-\frac{a}{2}}\right),\,\,\, \rho\gg\rho_1
\end{eqnarray}
The $a_i$ can be series expanded in $\alpha^2$ as
\begin{eqnarray}
\nonumber
a_0(\alpha)&=&\sum_{n=0}^{\infty}a_{0,n} \alpha^{2n}, \\
a_1(\alpha)&=&\sum_{n=0}^{\infty}a_{1,n}\alpha^{2n}.
\end{eqnarray}
In principle, we could numerically measure the $a_{i,n}$ by obtaining core solutions that obey our boundary conditions.  However, we elect to continue and leave our results in terms of the $a_{i,n}$.

The linear region is the domain where $y/\rho, y',A_{\psi},$ and $\rho A_{\psi}'\ll 1$ and so the quadratic Lagrangian Eq.~(\ref{abjmLinear}) is a good approximation to the full Lagrangian.  Unfortunately, we have not been able to decouple the fields $y$ and $A_{\psi}$ everywhere in this regime and thus have not found the linearly independent solutions for them.  However, we do know the solutions for them for $\rho\ll \sqrt{B}$ in the effective AdS$_2$.  The eigenmodes $\Phi_{\pm}$ fluctuate there as scalars of mass-squared $\mp B/\sqrt{\tilde{d}^2+B^2}$.  The effective AdS$_2$ region extends from $\rho$ of order but smaller than $\sqrt{B}$ to where our linearized approximations begin to fail.

For vanishing sources and small condensates we can further approximate the solutions in the linear region: we linearize in $c$ and $C$.  Near the transition the solutions in the effective AdS$_2$ region will then be
\begin{eqnarray}
\nonumber
\label{abjmLinPhi}
\Phi_+(\rho)&\sim&\rho^{-\frac{1}{2}}\left( \frac{2f_1}{\alpha}\sin\left(\frac{\alpha}{2}\log\rho\right)+f_2\cos\left(\frac{\alpha}{2}\log\rho\right) \right), \\
\Phi_-(\rho)&\sim&\rho^{-\frac{1}{2}}\left( f_3\rho^{\frac{a}{2}}+f_4\rho^{-\frac{a}{2}} \right),
\end{eqnarray}
where the $f_i$ are linear in $c$ and $C$ and depend upon $\alpha$.

The linear and core regimes overlap in the region $\rho\ll \sqrt{B}, \Phi_{\pm}, y',$ and $\rho A_{\psi}'\ll 1$.  This region is rather large near the transition and includes the infrared AdS$_2$ region.  We match the linear and core solutions at an arbitrary point in this region, which we take to be of order $\xi$, is near the bottom of the AdS$_2$.  Both the core the linear asymptotics overlap here so that we need only match the solutions Eqs.~(\ref{abjmCorePhi}) and (\ref{abjmLinPhi}).  In order for these to match the argument of the $\sin$ in Eq.~(\ref{abjmLinPhi}) must make $\pi$ between $\rho_1\sim \xi$ and $\rho\sim\sqrt{B}$.  To exponential accuracy we then have $\xi\sim e^{-2\pi/\alpha}$ which gives
\begin{equation}
c,C\sim f_i\sim  \sqrt{\xi}\sim e^{-\pi/\alpha}.
\end{equation}
Thus the chiral transition has exponential scaling and is of the holographic BKT type.

Next, the way that we matched the core and linear solutions makes it clear there are in fact an infinite number of symmetry-breaking embeddings.  Following~\cite{Jensen:2010ga} we term these solutions ``Efimov extrema'' for their resemblance to Efimov states~\cite{Efimov:1970zz}.  The most general match will have the argument of the $\sin$ in Eq.~(\ref{abjmLinPhi}) can pass through $n\pi$ between $\rho_1\sim\xi$ and $\rho\sim\sqrt{B}$ for $n$ any positive integer.  These embeddings oscillate through $n/2$ periods between the D2 branes at the bottom of AdS$_4$ and the boundary.  Matching the core and linear solutions at $\rho\sim\xi$ we find to exponential accuracy $\xi_n\sim e^{-2n\pi/\alpha}$, which yields condensates
\begin{equation}
c_n,C_n,\sim f_i\sim \sqrt{\xi_n}\sim e^{-n\pi/\alpha},
\end{equation}
where $c_1,C_1$ are the condensates for the simplest embedding above.  These embeddings, however, are not dual to ground states: denoting the free energy of the symmetric phase as $F_0$, we find that the free energy for the state dual to the $n^{th}$ non-trivial embedding goes like $F_0-F_n\sim e^{-2\pi n/\alpha}$.  It then follows that the $n=1$ embedding is dual to the ground state in the broken phase.  The extra non-trivial embeddings are then dual to extrema of the dual theory that are at best metastable.  The free energies of these extrema form an infinite tower, spaced by the same factor $e^{-2\pi/\alpha}$, reminiscent of the Efimov states.  These are an infinite exponential hierarchy of three-body bound states in atomic systems where two-body interactions have been tuned to threshold.

Finally, some words about finite temperature are in order.  We turn on a temperature in the field theory by heating up the D2 branes.  The dual geometry is a black brane geometry with an AdS$_4$-Schwarschild factor and a horizon radius $r_h\propto T$.  This change will dramatically affect the chiral transition.  In the D3/D5 system the chiral transition became second-order with mean-field exponents at any nonzero temperature~\cite{Jensen:2010ga,Evans:2010hi}.  Indeed, we expect the same thing to happen here.

The crux is that fluctuations on the brane no longer see an infrared AdS$_2$ geometry arbitrarily close to the horizon.  Rather, the equation of motion for fluctuations will have three separate distinct regimes: an AdS$_4$ region at large $r\l \sqrt{B}$, an infrared AdS$_2$ for $r_h\ll r\ll \sqrt{B}$, and a near-horizon geometry conformal to Rindler space for $r-r_h\ll r_h$.  The fields $\Phi_{\pm}$ are therefore stabilized in the deep infrared.  Their fluctuations can be described by an equivalent quantum mechanics problem with a non-trivial potential~\cite{Kaplan:2009kr}.  This potential has the form of an inverse square potential regulated at both short distances (by the control parameters) and at large distances (by the temperature).  As $\tilde{d}$ is decreased, the coefficient of the inverse square potential passes below the stable bound; however, no tachyons are formed because the potential is regulated in the IR and UV.  However, as $\tilde{d}$ is decreased further a single tachyon will eventually form signalling a second-order continuous transition rather than BKT scaling.  This is very different from the picture at zero temperature: there is no infrared regulator there and so an infinite exponential hierarchy of tachyons forms at the transition.  Thus the holographic BKT transition only occurs at zero temperature, as generally noted in~\cite{Jensen:2010ga,Iqbal:2010eh}.

The exponential scaling, however, is not utterly lost at nonzero temperature.  As shown in Fig.~\ref{cPlot} for the little string theory, the finite temperature condensates asymptote to the zero temperature result for large enough magnetic fields.  Moreover, the critical temperature will itself scale exponentially with $1/\alpha$~\cite{Iqbal:2010eh}.  The simple way to see this is to use the quantum mechanics picture for fluctuations.  For a scalar field in AdS$_{d+1}$ with $m^2=m_{\rm BF}^2-\alpha^2/4$ with ultraviolet and infrared regulators that become important at respective scales $r_{\rm UV}$ and $r_{\rm IR}$, the first tachyon forms when they are exponentially separated~\cite{Kaplan:2009kr}
\begin{equation}
\frac{\alpha}{2}\log\frac{r_{\rm UV}}{r_{\rm IR}}\sim \pi.
\end{equation}
The AdS$_2$ region is defomred in the infrared by the black brane horizon at $r_{\rm IR}\sim r_h$.  To exponential accuracy the finite temperature transition therefore occurs at
\begin{equation}
T_c\sim e^{-2\pi/\alpha}.
\end{equation}
These features should be quite general for holographic BKT transitions.

\subsection{Flavored little string theory}

This subsection is divided into three parts.  In the first, we do some analytic work and identify the BKT transition in this system.  Next we obtain some numerical results at zero temperature and find excellent agreement with our analytic predictions.  Finally, we look at this theory at finite temperature wherein we lose the BKT scaling near the transition.  Our discussion throughout will parallel our work above in the ABJM theory.

\subsubsection{Analytic results}

The onset of the chiral BKT transition can be found by analyzing the stability of fluctuations of the symmetric embedding $y=0$, given by the quadratic part of Eq.~(\ref{ldAction2})
\begin{equation}
\label{ldQuadratic}
\tilde{\mathcal{L}}_5\sim -\frac{\mathcal{N}\sqrt{\mu^2+\tilde{B}^2+\rho^2}}{2}y'^2 +\frac{\mathcal{N}\tilde{B}^2y^2}{2\rho^2\sqrt{\mu^2+\tilde{B^2}+\rho^2}}.
\end{equation}
We pause to consider the behavior of $y$ in two different limits.  At large $\rho\gg \mu,B,M$ the radial equation of motion for $y/\rho$ is that of a $m^2=-1$ scalar in AdS$_3$.  At small $\rho$ (the infrared of the dual theory), however, $y/\rho$ behaves like a $m^2=-\frac{\tilde{B}^2}{\mu^2+\tilde{B}^2}$ scalar in AdS$_2$.  This result is crucial: even though the D5 branes are embedded in the linear dilaton background at all $\rho$ the fields living on them see different geometries at different $\rho$.  The open string metric is dominated in the IR by the control parameters~\cite{Seiberg:1999vs}.

For $\mu^2/\tilde{B}^2 \geq 3$, $y/r$ fluctuates stably in the IR.  However, for $\mu^2/\tilde{B}^2<3$ the mass of $y/\rho$ in the IR is below the Breitenlohner-Freedman bound for stability~\cite{Breitenlohner:1982jf} in the effective AdS$_2$, $m^2_{\rm BF}=-1/4$.  The $y=0$ embedding is unstable in this phase and the stable phase will be $y\neq 0$.  There is therefore a chiral quantum phase transition at
\begin{equation}
\frac{\mu^2}{\tilde{B}^2}_c=3.
\end{equation}

We can compute thermodynamics in the broken phase by solving for the stable embeddings with appropriate boundary conditions.  Near the transition, we can find the embeddings by matching solutions in two regions: a small-$\rho$ non-linear core and a large-$\rho$ linear bulk.  We begin with the core region $\rho\ll \mu,B,M$ where the action looks like
\begin{equation}
\label{coreS}
\tilde{S}\sim -\mathcal{N}\int d\rho \sqrt{1+y'^2}\sqrt{\mu^2+\tilde{B}^2\frac{\rho^2}{r^2}}.
\end{equation}
Our brane embeddings minimize this action and obey the boundary condition $y(0)=0$.  Moreover, since this region enjoys a scaling symmetry we can find \emph{all} such embeddings: for $f(\rho)$ a particular solution that obeys the boundary condition, we define another solution $y_{\xi}(\rho)\equiv \xi f(\rho/\xi)$ for all $\xi\in \mathbb{R}_+$.  

At large $\rho$ (near the boundary of the IR AdS$_2$) $y/r$ has the asymptotics of a $m^2=-\frac{\tilde{B}^2}{\mu^2+\tilde{B}^2}$ scalar in AdS$_2$.  Since we are interested in the near-critical behavior, we will consider $\mu^2$ close to the critical value.  Defining 
\begin{equation}
\alpha=2\sqrt{m^2_{\rm BF}-m^2_{\rm IR}}=\sqrt{\frac{\mu_c^2-\mu^2}{\mu^2+\tilde{B}^2}},
\end{equation}
this region corresponds to $\alpha$ both real and small.  Then $f$ has the large-$\rho$ asymptotics
\begin{equation}
f(\rho)\sim\sqrt{\rho}\left( -a_0 \cos\left(\frac{\alpha}{2}\log \rho\right)+\frac{2a_1}{\alpha}\sin\left(\frac{\alpha}{2}\log\rho\right) \right),
\end{equation}
where $a_0$ and $a_1$ are functions of $\alpha$.  To order $\alpha$, a general embedding then has the large-$\rho$ asymptotics
\begin{equation}
\label{coreAsySol}
y(\rho)=\frac{2a_1}{\alpha}\sqrt{\xi\rho}\sin \left(\frac{\alpha}{2}\log\frac{\rho}{\rho_1}\right), \,\, \rho\gg \rho_1\equiv \xi e^{a_0/a_1}.
\end{equation}
Since we are interested in the near-critical embeddings, we series expand $a_0$ and $a_1$ in $\alpha$,
\begin{eqnarray}
\nonumber 
a_0(\alpha)&=&\sum_{n=0}^{\infty}a_{0,n}\alpha^{2n}, \\
a_1(\alpha)&=&\sum_{n=0}^{\infty}a_{1,n}\alpha^{2n}.
\end{eqnarray}
As we discuss in Sec.~\ref{ldNumerics}, we found solutions $f$ at various small $\alpha$ and numerically measured $a_{0,0}=-.32189$, $a_{0,2}=-2.328$, $a_{1,0}=.42525$, and $a_{1,2}=-.7477$.

The linear region is the domain where $y/\rho,y'\ll 1$ and the quadratic Lagrangian Eq.~(\ref{ldQuadratic}) is a good approximation to the full action.  The two linearly independent solutions for $y$ can be found here without any further approximation
\begin{equation}
\label{ldLinearSol}
f_{\pm}(u)=u^{\frac{1\pm i\alpha}{2}}\,_2F_1\left(\frac{1\pm i\alpha}{4},\frac{1\pm i\alpha}{4};1\pm\frac{i\alpha}{2};-u^2  \right),
\end{equation}
where
\begin{equation}
u\equiv \frac{\rho}{\sqrt{\mu^2+\tilde{B}^2}}.
\end{equation}
Since we are studying the theory at zero bare mass, we impose the large$-\rho$ boundary condition
\begin{equation}
\label{ldUVbc}
y(u)=c_+ f_+(u) + c_- f_-(u)=y_0+O\left(\frac{1}{u}\right),
\end{equation}
where we recall from Eq.~(\ref{ldVeVs}) that the condensate of the dual theory is simply $-\mathcal{N}y_0$.  Demanding the reality of the embedding forces $c_{\pm}$ to be complex conjugates, which together with the boundary condition Eq.~(\ref{ldUVbc}) determines
\begin{equation}
\label{cPM}
c_{\pm}=\frac{c}{\mathcal{N}}\left(\mp i\sum_{n=0}^{\infty}C_{2n+1}\alpha^{2n-1}+\sum_{n=0}^{\infty}C_{2n+2}\alpha^{2n}\right),
\end{equation}
where the $C_i$ are all real constants.  The first four are
\begin{eqnarray}
\nonumber
C_1=\frac{\sqrt{2}}{\pi}, \,\,\, \frac{C_2}{C_1}&=&-\frac{\log 8}{2}, \,\,\, \frac{C_3}{C_1}=-\frac{1}{2}\left(\frac{C_2}{C_1}\right)^2+\frac{\pi^2}{24}, \\
\frac{C_4}{C_1}&=&\frac{C_2C_3}{C_1^2}+\frac{1}{3}\left( \frac{C_2}{C_1} \right)^3-\frac{\zeta (3)}{4}.
\end{eqnarray}
At small $u$ near the IR AdS$_2$ boundary $y$ takes the form
\begin{equation}
\label{ldyMatch}
y(u)=\sqrt{u}(c_+u^{i\alpha/2}+c_-u^{-i\alpha/2}),
\end{equation}
which can be rewritten with Eq.~(\ref{cPM}) as
\begin{eqnarray}
y(u)=\frac{2C_1c\sqrt{u}}{\mathcal{N}\alpha}&&\text{\Huge{(}} \sum_{n=0}^{\infty}\frac{C_{2n+1}}{C_1}\alpha^{2n} \sin\left(\frac{\alpha}{2}\log u\right) \\ \nonumber
&&+\sum_{n=0}^{\infty}\frac{C_{2n+2}}{C_1}\alpha^{2n+1}\cos\left(\frac{\alpha}{2}\log u\right) \text{\Huge{)}}.\,\,
\end{eqnarray}
To order $\alpha$, these asymptotics can be rewritten as
\begin{equation}
\label{linAsy}
y(\rho)=\frac{2C_1c\sqrt{\rho}}{\mathcal{N}\alpha (\mu^2+\tilde{B}^2)^{1/4}}\sin\left( \frac{\alpha}{2}\log \frac{\rho}{\rho_2} \right),
\end{equation}
where
\begin{equation}
\rho_2=\sqrt{\mu^2+\tilde{B}^2}\exp \left(-\frac{2C_2}{C_1}\right)
\end{equation}
is an intrinsic length scale in the linear regime.

The linear and core regimes overlap in the region $\rho\ll \mu$, $y/\rho, y'\ll 1$.  This region is quite large near the transition.  We therefore perform the matching at an arbitrary point in this region, which we pick to be $\rho=\xi$ where the core and linear asymptotics overlap.  In order to match these solutions the argument of the $\sin$ in Eq.~(\ref{linAsy}) must make $\pi$ between $\rho_1\sim \xi$ and $\rho_2\sim \mu$.  To exponential accuracy we then have $\xi\sim e^{-2\pi/\alpha}$ which gives a condensate
\begin{equation}
c\sim \sqrt{\xi}\sim e^{-\pi/\alpha}.
\end{equation}
Thus the transition in the little string theory obeys BKT scaling.  The exponent and normalization of both $\xi$ and $c$ can be found by matching $y$ and $y'$ at $\xi$ order by order in $\alpha^2$
\begin{widetext}
\begin{eqnarray}
\nonumber
\label{cPredict}
\xi &=& \sqrt{\mu^2+\tilde{B}^2}\,\exp \left( -\frac{2\pi}{\alpha}-\frac{a_0}{a_1}-\frac{2C_2}{C_1}+\alpha^2\left(\underbrace{ \frac{a_{00}^3}{12 a_{10}^3}+\frac{2}{3C_1^3}(C_2^3+3C_1C_2C_3-3C_1^2C_4)}_{\equiv \xi_2} \right)+O(\alpha^4) \right), \\
c&=&-\mathcal{N}\sqrt{\mu^2+\tilde{B}^2}\,\frac{a_{1}}{C_1}\exp\left( -\frac{\pi}{\alpha}-\frac{a_0}{2a_1}-\frac{C_2}{C_1}+\alpha^2\left( \frac{\xi_2}{2}+\frac{a_{00}^2}{8a_{10}^2}-\frac{1}{2C_1^2}(C_2^2+2C_1C_3) \right)+O(\alpha^4) \right).
\end{eqnarray}
\end{widetext}
  We also compare this result to numerical data in Fig.~\ref{cPlot}.

As before, the way that we matched the core and linear solutions makes it clear there are an infinite number of embeddings which, following~\cite{Jensen:2010ga} we again term ``Efimov extrema'' for their resemblance to Efimov states~\cite{Efimov:1970zz}.  Indeed, the most general match has the argument of the $\sin$ in Eq.~(\ref{linAsy}) pass through $n\pi$ between $\rho_1\sim\xi$ and $\rho_2\sim \mu$, for $n$ a positive integer.  These embeddings go through $n/2$ oscillations between the bottom of the brane and the boundary.  Matching $y$ and $y'$ at $\rho=\xi$ we find a scale factor and condensate corresponding to these extrema of
\begin{eqnarray}
\nonumber
\xi_n &=&e^{-\frac{2(n-1)\pi}{\alpha}}\xi_1+O(\alpha^4), \\
c_n&=& (-1)^{n+1}e^{-\frac{(n-1)\pi}{\alpha}}c_1+O(\alpha^4),
\end{eqnarray}
where $\xi_1$ and $c_1$ are the scale factor and condensate for the simplest embedding above.  However, the additional extrema are not ground states: if one denotes by $F_0$ the free energy corresponding to the symmetric embedding, then the free energy of the $n^{\rm th}$ non-trivial extremum goes like $F_0-F_n\sim e^{-2\pi n/\alpha}$, so the $n=1$ embedding is the ground state.  The infinite tower of extrema, spaced by the same factor $e^{-2\pi/\alpha}$ is reminiscent of the Efimov states.

\subsubsection{Numerical results}
\label{ldNumerics}

We have two tasks in this subsection.  The first is to identify how we numerically obtain core solutions (i.e. solutions that minimize Eq.~(\ref{coreS})).  The second is to numerically obtain full zero-temperature numerical solutions that minimize the whole action Eq.~(\ref{ldAction2}).  We will obtain numerical solutions at finite temperature in Sec.~\ref{finiteT}.

\begin{figure}[t]
\includegraphics[scale=0.8]{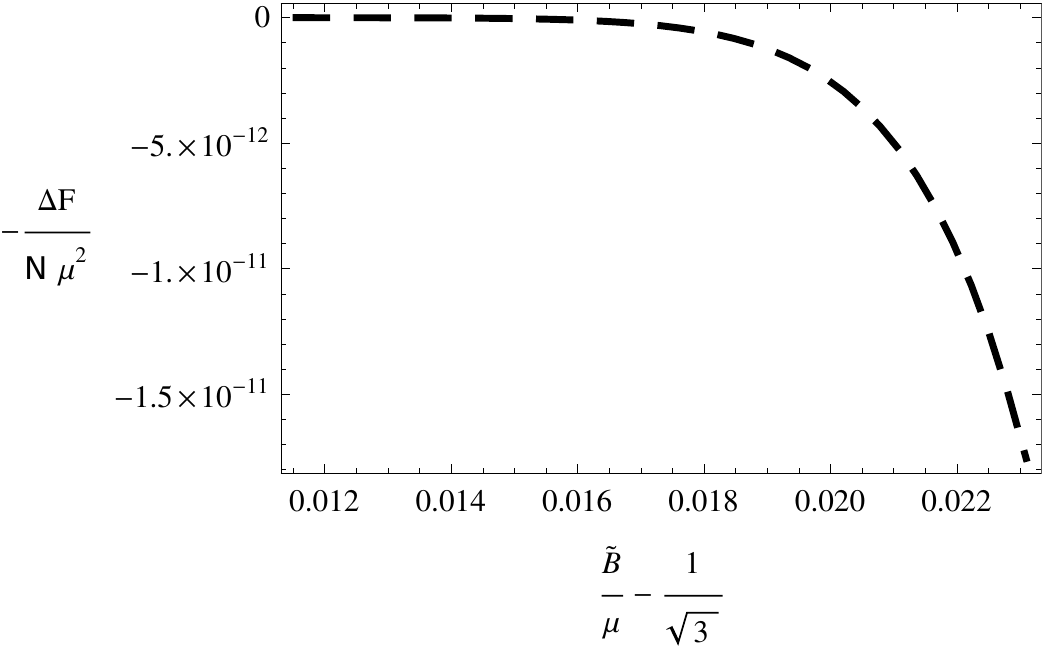}
\caption{\label{fPlot}
The free energy difference between the broken and symmetric phases of the little string theory as a function of magnetic field.  This curve scales exponentially as $\Delta F\sim e^{-2\pi/\alpha}$.
}
\end{figure}
We find numerical core solutions by shooting from small $\rho$, i.e. at the very bottom of the brane.  In order to find a good series solution here we elect to change coordinates and redefine our worldvolume fields by
\begin{equation}
y=e^{-\tau}\cos \phi, \,\,\, \rho=e^{-\tau}\sin \phi.
\end{equation}
The core action in these coordinates is
\begin{equation}
\tilde{S}_5\sim -\mathcal{N}\int d\tau e^{-\tau}\sqrt{1+\phi'^2}\sqrt{\mu^2+\tilde{B}^2\sin^2\phi}.
\end{equation}
As the reader may note from the form of the behavior of $y$ in the AdS$_2$ region Eq.~(\ref{ldyMatch}), both $y/\rho$ and $y'$ blow up at the bottom of the brane.  This corresponds to $\phi=\phi(\tau)$ going to zero as $\tau\rightarrow\infty$.  Indeed, there is a one-parameter family of series solutions that obey this boundary condition,
\begin{equation}
\label{coreSeries}
\phi(\tau)=\phi_0 e^{-m\tau}\left(1+\sum_{n=1}^{\infty}\phi_i e^{-2m\tau}\right),
\end{equation}
where
\begin{equation}
m=\frac{-1+\sqrt{1+4\frac{\tilde{B}^2}{\mu^2}}}{2},
\end{equation}
and the $\phi_i$ can be recursively solved for in terms of $\phi_0$.  Since this series expansion is in powers of $e^{-m\tau}$ we could write a series solution for $\phi$ in terms of the radial coordinate $r$ as $\phi=\phi_0 r^{m}+O(r^{3m})$.  In any event, we use this small$-\rho$ series solution as initial data upon which we numerically integrate to find $y$ up to very large $\rho$ close to the AdS$_2$ boundary (of order $\rho\sim 10^{25}$).  We then match the numerical solution to the near-AdS$_2$ boundary solution Eq.~(\ref{coreAsySol}) to find $a_0$ and $a_1$ at small $\alpha$.

\begin{figure}[t]
\includegraphics[scale=0.8]{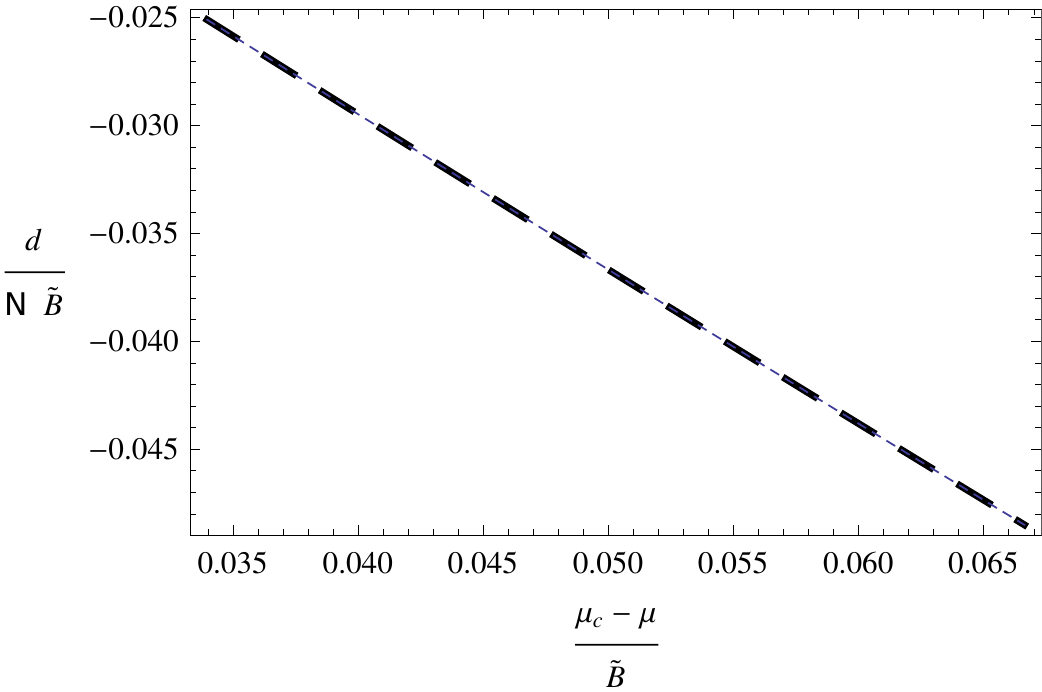}
\caption{\label{muVSd}The density as a function of chemical potential in the broken phase of the little string theory.  The thin dashed blue curve represents a direct measurement of the density by Eq.~(\ref{ldVeVs}) while the thick dashed black curve is $-\partial F/\partial \mu$.}
\end{figure}

Our full zero-temperature solutions were obtained by shooting from large $\rho$.  Imposing the zero mass condition, we find a series solution for $y$ near infinity
\begin{equation}
y(\rho)=y_0+\sum_{n=1}^{\infty}y_{n}\rho^{-2n},
\end{equation}
where the $y_n$ are recursively determined by $y_0$.  We use this series solution as initial data upon which we again numerically integrate.  This time we proceed from large $\rho$ (here, we use a series solution to eight orders and integrate from $\rho=100$) to very small $\rho$ (of order $\rho\sim 10^{-30}$ to $10^{-28}$).  However, due to the spiky behavior of the core solution Eq.~(\ref{coreSeries}) near $\rho=0$ we need to more careful in order to ensure that we correctly shoot to the IR boundary condition $y(0)=0$. This simply means that we shoot for our solutions match the core series solution Eq.~(\ref{coreSeries}) at small $\rho$.

We plotted some of these results in Sec.~\ref{intro}.  First, we plotted the condensate in the broken phase in Fig.~\ref{cPlot}.  We also computed the regularized free energy $F=-S_{5,\rm ren}$ in the broken phase and plotted it in Fig.~\ref{fPlot}.  The agreement between our analytical and numerical data for the condensate indeed holds to $O(\alpha^4)$.

Finally, we recall our prescription for computing the density $d$ in the dual theory.  In principle we can also obtain the density from the free energy: thermodynamic consistency requires that $d=-\partial F/\partial \mu$.  We compare these in Fig.~\ref{muVSd} after using Eq.~(\ref{ldVeVs}) and (\ref{ldDensity}) to measure $d$.  Indeed the two agree remarkably well - our prescription for the free energy of the dual theory passes this consistency test at zero temperature.

\subsubsection{Finite temperature}
\label{finiteT}

\begin{figure}[t]
\includegraphics[scale=0.8]{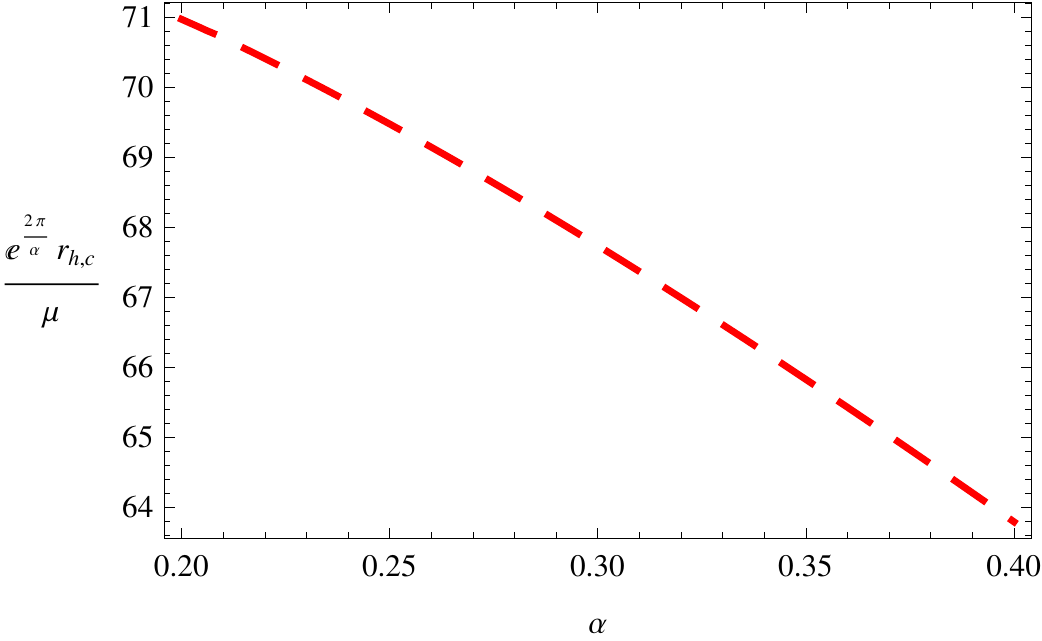}
\caption{\label{critT}
The rescaled critical horizon radius as a function of $\alpha$ in the little string theory.  Over this domain $e^{2\pi/\alpha}$ decreases by a factor $\sim6.64\times 10^6$, while the rescaled radius is relatively constant.  The critical radius therefore scales as $r_{h,c}\sim e^{-2\pi/\alpha}$.
}
\end{figure}

The action Eq.~(\ref{ldAction2}) is only suitable at zero temperature.  We turn on a temperature in the little string theory by heating up the 5-branes.  The resulting geometry is a black brane geometry
\begin{eqnarray}
\nonumber g&=&\frac{r}{R}(-f dt^2+d\vec{x}^2)+\frac{dr^2}{rRf}+rRg_{\mathbb{S}^3}, \\
&&\hspace{.8cm}f(r)=1-\frac{r_h^2}{r^2},
\end{eqnarray}
where $r_h$ is the horizon radius and the dilaton and form fields are as before.  This black brane has some interesting thermodynamics:  its temperature is independent of $r_h$,
\begin{equation}
T=\frac{1}{2\pi R}.
\end{equation}
This fact informs us that (i.) the little string theory has a Hagedorn phase and (ii.) the bulk theory at any (order $1$) temperature is dual to that Hagedorn phase.  Consistent with this result, the energy and entropy densities of the black brane grow with $r_h$,
\begin{equation}
s=\frac{r_h^2}{(2\pi)^4g_s^2R}.
\end{equation}
In both the bulk and field theories, entropy and energy can be added by increasing $r_h$ without raising the temperature.

We will now measure how the BKT transition is modified in the presence of this background entropy density.  We will use a different parametrization for the embedding than in Eq.~(\ref{ldZeroTMetric}), writing the 3-sphere metric as
\begin{equation}
g_{\mathbb{S}^3}=d\theta^2+\sin^2\theta d\phi^2+\cos^2\theta d\psi^2,
\end{equation}
and letting $\theta=\theta(r)$ depend on the radial coordinate.  As before our brane embeddings will be specified by $\phi=x^4=x^5=$constant.  The Legendre transformed brane action density now takes the form
\begin{equation}
\tilde{S}_5=-\mathcal{N}\int dr\sqrt{1+r^2f\theta'^2}\sqrt{\mu^2+r^2\cos^2\theta\left(1+\frac{\tilde{B}^2}{r^2}  \right)}.
\end{equation}
Before going on, we should pause to note an important result.  The linearized action for small fluctuations around the chirally symmetric $\theta=0$ is now
\begin{equation}
\mathcal{L}\sim -\mathcal{N}\sqrt{\mu^2+\tilde{B}^2+r^2}\frac{r^2f\theta'^2}{2}+\frac{\mathcal{N}(\tilde{B}^2+r^2)\theta^2}{2\sqrt{\mu^2+\tilde{B}^2+r^2}},
\end{equation}
which, as for the ABJM theory, now exhibits three separate limits at small horizon radius.  At large $r\gg \mu$, $\theta$ fluctuates like a stable scalar field and there is the AdS$_2$ region at medium $r_h\ll r\ll \mu$.  However, there is now a near-horizon region $r-r_h\ll r_h$, conformal to Rindler space, where the field is stable.  Following the same approach as in Sec.~\ref{abjmResults}, we describe this system with an equivalent quantum mechanics.  This problem has a potential that has the form of an inverse square potential regulated at both small and large distances.  As before a single tachyon will form as $\mu$ is decreased, signalling a continuous transition rather than exponential scaling.

Moreover we can estimate the scaling of the critical horizon radius at fixed $\alpha$~\cite{Iqbal:2010eh}.  Using the same logic as in Sec.~\ref{abjmResults}, we find that to exponential accuracy the finite temperature transition occurs at
\begin{equation}
r_{h,c}\sim e^{-2\pi/\alpha}.
\end{equation}

We confirm these results numerically by solving the embedding equations at nonzero temperature.  We do so again by shooting inwards from the boundary.  Our boundary condition there is that $\theta$ is normalizable, for which we obtain a large$-\rho$ series solution
\begin{equation}
\theta(r)=\sum_{n=0}^{\infty} \frac{\theta_{2n}}{r^{2n+1}},
\end{equation}
where the $\theta_n$ are recursively determined by $\theta_0$.  We use this series solution to set initial conditions for numerical integration.  We then integrate down very close to the horizon, where we see if our solution satisfies the infrared boundary condition that $\theta$ is regular at $r_h$.

\begin{figure}[t]
\includegraphics[scale=0.8]{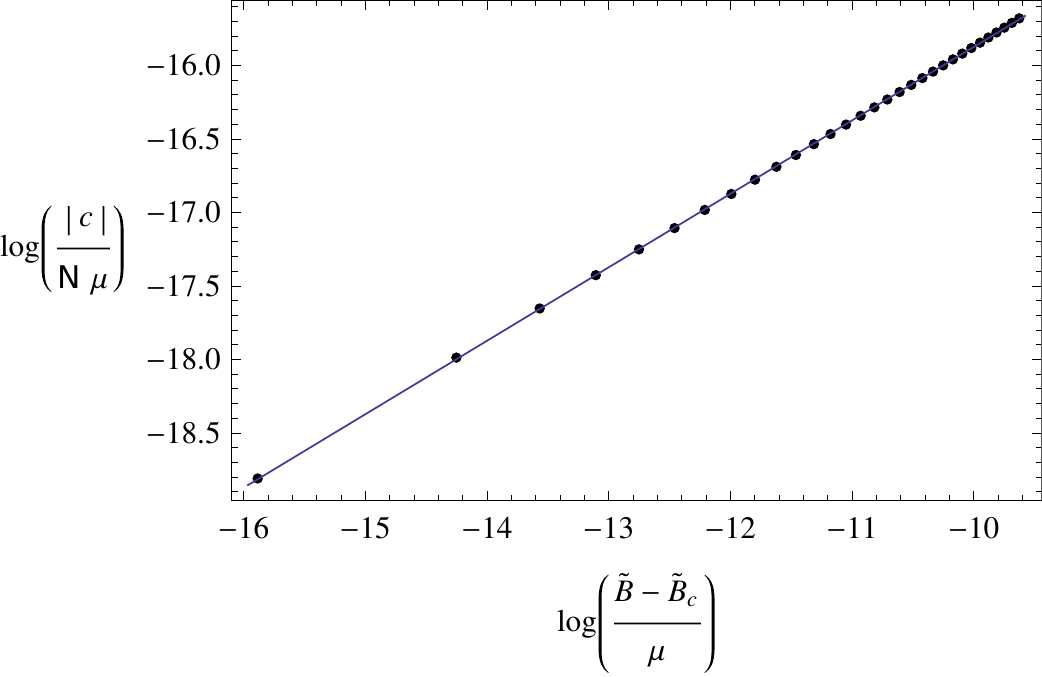}
\caption{\label{cLog}
A log-log plot of the condensate as a function of magnetic field at nonzero temperature in the little string theory.  The black dots are numerical data and the thin blue line is a fit with slope $.500\sim1/2$.  This slope measures the critical exponent $\nu$, which takes the mean-field value.
}
\end{figure}

We plotted some of our results in Sec.~\ref{intro}.  We plotted the condensate in Fig.~\ref{cPlot}; unfortunately, we have suffered some systematic errors in our numerics for the free energy at nonzero temperature and so we do not plot them here\footnote{Away from the chiral transition at nonzero temperature the embeddings asymptote to the zero temperature result for $r-r_h\gg r_h$.  For small horizon radii the free energy should then also asymptote to the zero temperature data.  We do not find this, implying some systematic errors in our determination of the free energy at finite temperature.}.  We go on to present more data in this subsection.  First, we  plot the critical horizon radius as a function of $\alpha$ and find that it indeed scales exponentially in Fig.~\ref{critT}.  Finally, we measure the static critical exponent $\nu$,
\begin{equation}
|c|\sim (\mu_c(T)-\mu)^{\nu},
\end{equation}
for the transition at $r_h=10^{-12}$ in Fig.~\ref{cLog} and find it to be mean-field, $\nu=0.500\sim 1/2$.

\section{Discussion}
\label{discuss}
\subsection{IR AdS$_2$}
\label{irAdS2}
We have identified quantum holographic BKT transitions in both flavored ABJM and little string theories.  The mechanism in the bulk that drives the transition is clear: in each case the field dual to the chiral symmetry breaking order parameter becomes unstable near the bottom of the flavor brane.  The radial equation for the field in this region is that of a scalar in AdS$_2$ with a mass that depends on the control parameters.  The holographic BKT transition occurs when the mass drops below the Breihtenlohner bound for stability~\cite{Breitenlohner:1982jf} in this effective AdS$_2$ region.

We would like to use this information to infer properties of the mechanism in the field theory that gives rise to the holographic BKT transition.  As in~\cite{Jensen:2010ga} we find it instructive to consider a multiplicity of systems that do \emph{not} realize holographic BKT transitions.  The flavored little string theory, in particular, is one of many systems whose holographic dual has supersymmetric flavor branes probing the near-horizon $p$-brane geometries~\cite{Itzhaki:1998dd}.  These geometries are given by
\begin{eqnarray}
\nonumber
g&=&Z(r)^{\frac{7-p}{4} }g_{p,1}+Z(r)^{\frac{p-7}{4}}(dr^2+r^2g_{\mathbb{S}^{8-p}}), \\
e^{\phi}&\propto&Z(r)^{\frac{p-3}{4}}, \,\,\, Z(r)=\frac{r^2}{R^2},
\end{eqnarray}
where $R$ is related to the coupling of the Yang-Mills theory on the $p$-branes and there are $N$ units of $(p+2)$-form flux.  For $p<7$ these field theories typically flow to flavored maximally supersymmetric Yang-Mills theories in limits that vary with $p$.  Most supersymmetric probe branes in these backgrounds fall into one of three categories: (i.) D$(p+4)$ branes that wrap an $\mathbb{S}^3\subset\mathbb{S}^{8-p}$, dual to codimension-0 flavor, (ii.) D$(p+2)$ branes that wrap an $\mathbb{S}^2$, dual to codimension-1 flavor, and (iii.) D$p$ branes that wrap an $\mathbb{S}^1$, dual to codimension-2 flavor.  The others have flavor in either $(0+1)$ or $(1+1)$ dimensions, for which we cannot turn on a magnetic field.

The probe brane action for the theory with condimension-$k$ flavor at nonzero magnetic field and density is not hard to obtain.  Following the derivation of Eq.~(\ref{ldAction}), we find that it is proportional to
\begin{equation}
\tilde{S}_{p,k}\propto \int d^{p+1-k}xd\rho \sqrt{1+y'^2}\sqrt{d^2+\rho^{2(3-k)}\left(1+\frac{B^2}{r^{7-p}}\right)},
\end{equation}
where $d$ is proportional to the constant of integration associated with $A_0'$ in the DBI action, $\partial \mathcal{L}/\partial A_0'$.  For most cases this constant is interpreted as the baryon density of the dual theory.  For others (like the flavored little string theory we consider), it is the chemical potential for the baryon density.  Also, $p+1-k$ must be greater than or equal to $3$ so that we can add a magnetic field.

The Lagrangian for embeddings close to the chirally symmetric one $y=0$ is
\begin{eqnarray}
\nonumber
\mathcal{L}&\propto& -\sqrt{d^2+B^2\rho^{p-2k-1}+\rho^{2(3-k)}}\frac{y'^2}{2} \\
&& +\frac{B^2\rho^{p-2k-3}}{\sqrt{d^2+B^2\rho^{p-2k-1}+\rho^{2(3-k)}}}\frac{(7-p)y^2}{4}.
\end{eqnarray}
This Lagrangian has an infrared limit that depends on whether $p-2k-1$ is positive, negative, or zero.  They have the case structure 
\begin{enumerate}
\item  $p-2k-1<0$: There is one such case, $p=4$ and $k=2$.  It has a limit where $y/\rho$ obeys the equation of motion of a $m^2=-1$ scalar in AdS$_{\frac{3}{2}}$ and so is unstable - the symmetric embedding is unstable at all densities.

\item $p-2k-1=0$: There are two cases, $p=3,k=1$ and $p=5,k=2$.  These are respectively the D3/D5 system and the flavored little string theory we consider in this work.  In these cases $y/\rho$ obeys the equation of motion in the infrared of a $m^2=-(3-k)B^2/(B^2+d^2)$ scalar in AdS$_2$, which passes below the BF bound for $d \leq \sqrt{7}B$ in the D3/D5 system and $d\leq \sqrt{3}B$ in the little string theory.  In each case there is a holographic BKT transition at this magnetic field.

\item $p-2k-1>0$: In all of these systems, $y/\rho$ behaves in the infrared as massless scalar in AdS$_2$. A large enough magnetic field will break chiral symmetry but the transition will not be of the holographic BKT type.  From the numerical work done in the D3/D7 system~\cite{Jensen:2010vd} we expect the transition to be second-order with mean-field exponents.   The one exception is $p=6$, $k=0$ where the magnetic field does not trigger chiral symmetry breaking at zero density; presumably there is no transition in that system at any magnetic field.
\end{enumerate}

The only theories in this class that realize holographic BKT transitions in the $(B,d)$ plane are the D3/D5 and D5/D5 systems.  We would like to understand what is special about these theories.  In~\cite{Jensen:2010ga}, we speculated that the special feature was that the theory lived in $(2+1)$-dimensions so that $B$ and $d$ had the same operator dimension.  The D5/D5 system is a striking counterexample, but can be understood similarly in terms of generalized conformal symmetry.

\subsection{Generalized conformal symmetry}

Most of these field theories run in the infrared (excepting flavored $\mathcal{N}=4$ SYM).  However, it is clear from their holographic duals that a \emph{generalized conformal symmetry} emerges there~\cite{Kanitscheider:2009as}.  Dilatation in the boundary theory is realized in the bulk by simultaneously scaling the field theory directions $x$ and the radial coordinate.  For a probe brane that adds codimension-$k$ flavor, the volume of the brane
\begin{equation}
V_k=\int d^{p+5-2k}\xi e^{-\phi}\sqrt{-\text{P}[g]}
\end{equation}
is invariant under
\begin{equation}
\mathcal{D}_{\lambda}: \,\, x\mapsto\lambda^{-1} x, \,\,\, r\mapsto\lambda^{\frac{4(p+1-k)}{1-4k+p(8-p)}}r.
\end{equation}
There is then a \emph{generalized} dilatation invariance in the dual flavor, under which operators will generally assume a dimension that differs from its canonical one.  As an example, let us obtain the dimensions of the magnetic field and chemical potential/density in these theories.

To find these dimensions we need only consider the $U(1)$ gauge field in the weak field limit,
\begin{equation}
S_{\rm EM}=-N_f T_{p+4-2k}\int d^{p+5-2k}\xi e^{-\phi}\sqrt{-\text{P}[g]}\frac{F^2}{4}.
\end{equation}
This action is invariant under $\mathcal{D}_{\lambda}$ when the components of the field strength assume particular eigenvalues under $\mathcal{D}_{\lambda}$.  For example, $F_{0r}$ must have dimension $0$ and $F_{ij}$ dimension $2(p+1-k)(7-p)/(1-4k+p(8-p))$.  From this we learn the dimensions
\begin{eqnarray}
\nonumber
[\mu ]&=&\frac{4(p+1-k)}{1-4k+p(8-p)}, \,\,\, [d]=(3-k)[\mu], \,\,\,  \\
&&\hspace{.8cm}[B]=\frac{2(p+1-k)(7-p)}{1-4k+p(8-p)},
\end{eqnarray}
where $[\mu ]$ comes from the scaling of the leading term of $A_0$ and $[d]$ from the near-boundary expansion of $A_0=\mu-d/\mathcal{N}(2-k)\rho^{2-k}+...$.  These dimensions obviously differ from the canonical ones $[\mu ]=1, [B]=2$.

Now we note that the only $(p,k)$ for which $[d]=[B]$ are $(3,1)$ and $(5,2)$.  These are the D3/D5 system and the flavored little string theory we study in this work.  In each case $B$ and $d$ have dimension $2$.  Indeed, from our analysis in Sec.~\ref{irAdS2} we know that these are the only systems in this class that realize holographic BKT transitions.  

What about the other $(p,k)$ systems?  It turns out that the one system with $p-2k-1<0$, $(4,2)$, has $[B]>[d]=[\mu]$.  The reader will recall that there is no chiral phase transition in this theory; chiral symmtrey is broken for all $B\neq 0$.  Moreover, the other systems with $p-2k-1>0$ have $[B]<[d]$.

These results are extremely suggestive in terms of renormalization group flow.  Consider a (generalized) conformal theory deformed with some set of control parameters.  These will induce flows of conjugate operators.  In particular, the infrared of the theory will be dominated by the flow of the lowest dimension operator which will be induced by the highest dimension control parameter.  If there are two or more control parameters of the same dimension are turned on, they can dominate the infrared, provided that they remain the same dimension (relative to each other) after being turned on.  In such a theory there will be non-trivial renormalization group flow as these operators compete to arbitrarily low energy scales.  We claim that these conditions are sufficient to engineer a holographic BKT transition if one of these operators tends to restore a symmetry and the other to break it.  At the least, this sort of scenario is markedly different than the usual Ginzburg-Landau-Wilson picture for phase transitions~\cite{SVBSF}.

This picture is certainly consistent with all known quantum critical points embedded in string theory.  The three probe brane systems with $[B]=[d]$ realize holographic BKT transitions.  The theory with $[B]>[d]=[\mu]$ has an infrared dominated by the magnetic field and there is never a chiral transition.  In the theories where $[B]<[d]$, the infrared is density-dominated and the chiral transitions are presumably second-order.

We have two loose ends to discuss.  The first is the scalar mass operator in the flavored little string theory.  How does it fit into this picture of generalized conformal invariance and BKT transitions?  A simple computation shows that $[M]=2$ in this theory, so that all three control parameters $\mu, B$, and $M$ have the same generalized conformal dimension.  We are not surprised, then, to find that there is a line of BKT transitions in the $(\mu,B,M)$ plane.  Second, there are two more systems for which $[\mu]=[B]$: $(p,k)=(5,0)$ and $(5,1)$.  Why do we not see holographic BKT transitions in these systems?  Our answer is simple.  In each case the dimension of the density is larger than $[\mu]$.  As a result the deep infrared will be dominated by the density, rather than an interplay between the magnetic field and $\mu$.  

\acknowledgements
It is a pleasure to thank Andreas Karch and Dam T. Son for their insight, comments, and suggestions.  The author would also like to thank Carlos Hoyos, Ethan Thompson, and Steve Paik for many useful comments and discussions.   This work was supported, in part, by the U.S. Department of Energy under Grant Numbers DE-FG02-96ER4095.

\appendix
\section{Computation of the spectrum}
\label{appSpectrum}
We obtain the meson spectrum of the flavored ABJM theory by studying small fluctuations around the embedding $y=m$.  There are four types of fluctuations corresponding to the four types of fields on the flavor brane, one corresponding to the transverse scalars $y$ and those related by $SO(3)_{\chi}$ symmetry and three to the gauge field on the brane.  We begin with a fluctuation of the transverse scalar $y=m+\delta y$.

The linearized Lagrangian for $\delta y$ is equivalent to
\begin{eqnarray}
\nonumber
\mathcal{L}_{\delta y}&&\propto \sqrt{-\text{P}[g]}\left[\delta y'^2+\frac{\partial_{\mu}y\partial^{\mu}y}{(\rho^2+m^2)^2}\right] \\
&&+\frac{2\rho}{\sqrt{\rho^2+m^2}}\sqrt{g_{\mathcal{C}_3}}\partial_i y\partial^iy,
\end{eqnarray}
where $\text{P}[g]$ is the induced metric on the background $y=m$, $\partial_i$ is a derivative with respect to a coordinate on $\mathcal{C}_3$ and the $i$'s are contracted with the metric on $\mathcal{C}_3$.  By separability, the equation of motion for $\delta y$ can then be solved by an eigenfunction expansion.  They are given by
\begin{equation}
\label{deltaY1}
\delta y_{k,n,l}=e^{i k\cdot x}\phi_{y,n}(\rho)\mathcal{Y}^j(\mathcal{C}_3),
\end{equation}
where the $\mathcal{Y}^j(\mathcal{C}_3)$ are harmonics on $\mathcal{C}_3$,
\begin{equation}
D_i D^i \mathcal{Y}^{l}(\mathcal{C}_3)=-\lambda_j \mathcal{Y}^j(\mathcal{C}_3),
\end{equation}
and $D_i$ is the covariant derivative on $\mathcal{C}_3$.  Since at any constant $\rho$ $\mathcal{C}_3$ is a squashed $\mathbb{RP}^3$ we can obtain the harmonics on $\mathcal{C}_3$ from the harmonics on $\mathbb{RP}^3$.  These fall into $\left(\frac{j}{2},\frac{j}{2}\right)$ multiplets of the $SO(3)\times SO(3)$ isometry of $\mathbb{RP}^3$ for integer $j/2$.  Under the pullback from the squashed $\mathbb{RP}^3$ to the round one, these harmonics form a complete set on the squashed space.  However, the reduced isometry of the squashed $\mathbb{RP}^3$ implies that these harmonics can mix and that their eigenvalues may be dependent upon the squashing parameter $y^2/\rho^2$.

Actually, only the second $SO(3)$ is broken to $U(1)$ in the squashed space.  The harmonics on $\mathcal{C}_3$ therefore fall into $\left(\frac{j}{2};\frac{j}{2},\frac{m'}{2}\right)$\footnote{This notation indicates both of the $SO(3)$ quantum numbers as well as the $U(1)$ charge.} representations of $SO(3)\times U(1)$ for integer $j/2$ and $m'/2$.  There is only one such representation for each $j/2$, $m'/2$, so the scalar harmonics do not mix on $\mathcal{C}_3$.  It is a short exercise to show that, however, the eigenvalues of the $\left(\frac{j}{2};\frac{j}{2},\frac{m'}{2}\right)$ harmonics are modified to depend on $y^2/\rho^2$ through
\begin{equation}
\label{scalarLambda}
\lambda_{\j,m'}=m'^2+\frac{j(j+2)-m'^2}{1+\frac{y^2}{\rho^2}}.
\end{equation}
The equation of motion for a mode Eq.~(\ref{deltaY1}) is then
\begin{equation}
\frac{(\rho\sqrt{\rho^2+m^2}\phi_y')'}{\rho\sqrt{\rho^2+m^2}}-\frac{ k^2}{(\rho^2+m^2)^{2}}\phi_y - \frac{1}{4\rho^2}\lambda_{j,m'}\phi_y = 0,
\end{equation}
which has the solutions
\begin{eqnarray}
\phi_{y,\pm}&&=\rho^{\pm \frac{m'}{2}}(\rho^2+m^2)^{4\beta} \\
\nonumber
&&\times _2 F_1\left( \frac{-j\pm m'+\beta}{4},\frac{2+j\pm m'+\beta}{4};1\pm \frac{m'}{2};-\frac{\rho^2}{m^2}\right),
\end{eqnarray}
where we have defined $\beta=1+\sqrt{1-4k^2/m^2}$.  We obtain the meson masses from these solutions by finding those $-k^2$ for which we satisfy boundary conditions at small and large $\rho$.  Near the bottom of the brane at small $\rho$ the $\phi_{y,\pm}$ behave as
\begin{equation}
\phi_{y,\pm}\sim \rho^{\pm \frac{m'}{2}}.
\end{equation}
Our small-$\rho$ boundary condition is simply that $\delta y$ is regular so that the correct solution is $\phi_{y,\text{sgn}(m')}$.  At large $\rho$ where we have
\begin{equation}
\label{deltaYlargeR}
\phi_{y,\pm}\sim \frac{\Gamma\left(\frac{j+1}{2}\right)\Gamma\left(1\pm \frac{m'}{2}\right)}{\Gamma\left( \frac{4+j\pm m'-\beta}{4} \right)\Gamma\left(\frac{2+j\pm m'+\beta}{4}  \right)}\rho^{\frac{j}{2}}+O\left( \rho \right)^{-1-\frac{j}{2}},
\end{equation}
the proper boundary condition is that $\phi_y$ is normalizable.  That is, the coefficient on the $\rho^{j/2}$ term must vanish.  This can only be accomplished for particular $k^2$ so that the denominator is sitting on a pole.  These are given by
\begin{equation}
\label{yMesons}
-k^2_y=\frac{m^2}{4}(2+j+|m'|+4n)(4+j+|m'|+4n).
\end{equation}
These are the meson masses(-squared) of the dual theory.  Moreover, the large-$\rho$ behavior of the solutions Eq.~(\ref{deltaYlargeR}) yields the dimensions of the dual operators, $\Delta = 2+\frac{j}{2}$.  These operators are of the form $\bar{\psi}(A_iB_i)^{j/2}\psi$, where the $\psi$ are the flavored fermions and the $A_i$'s and $B_i$'s are bifundamental fields of the ABJM theory~\cite{Hikida:2009tp}.

We move on to consider the fluctuations of the gauge fields.  Following~\cite{Kruczenski:2003be,Hikida:2009tp}, these can be broken into three classes,
\begin{widetext}
\begin{eqnarray}
\nonumber
\text{Type I:}&&\hspace{.5cm} A_{\mu}=\zeta_{\mu}e^{ik\cdot x}\phi_{I,n}(\rho)\mathcal{Y}^j(\mathcal{C}_3), \,\,\,\, A_{\rho}=0, \,\,\,\, A_i=0, \\
\nonumber
\text{Type II:}&& \hspace{.5cm}A_{\mu}=0, \,\,\,\, A_{\rho}=e^{i k\cdot x} \phi_{II,n}(\rho)\mathcal{Y}^j(\mathcal{C}_3), \,\,\,\, A_i=e^{ik\cdot x}\tilde{\phi}_{II,n}(\rho)D_i\mathcal{Y}^j(\mathcal{C}_3), \\
\text{Type III:} && \hspace{.5cm} A_{\mu}=0, \,\,\,\, A_{\rho}=0, \,\,\,\, A_i=e^{ik\cdot x}\phi_{III,n}(\rho) \mathcal{Y}_{i}^j(\mathcal{C}_3),
\end{eqnarray}
\end{widetext}
where we have indexed the internal directions with $i$ and the vector harmonics of spin $j/2$ under the unbroken $SO(3)$ on $\mathcal{C}_3$ are labeled by $\mathcal{Y}_{i}^j(\mathcal{C}_3)$.  These obey
\begin{eqnarray}
\nonumber
&&D_iD^i\mathcal{Y}_{k}^j(\mathcal{C}_3)-R^i_k \mathcal{Y}_{i}^j(\mathcal{C}_3)=-\lambda_j \mathcal{Y}_{k}^j(\mathcal{C}_3), \\
&&D^i\mathcal{Y}_{i}^j=0,
\end{eqnarray}
where $[D_i,D_k]=R_{ik}$ is the Ricci curvature.

We have been able to solve the meson spectrum for the Type I fluctuations.  Due to the background Ramond-Ramond 3-form potential the gauge field obeys
\begin{equation}
\label{eomA}
\partial_a(\sqrt{-\text{P}[g]} F^{ab})-\frac{3}{2}r^3\sin\theta \epsilon^{bij}\partial_i A_j=0,
\end{equation}
where we have chosen the orientation $(\psi,\theta,\phi)$ to be positive.  For the Type I fluctuations the $b=\mu$ equations impose the usual condition $\zeta\cdot k=0$, the $b=i$ equations are automatically satisfied, and the $b=\rho$ equation is
\begin{equation}
\frac{(\rho\sqrt{\rho^2+m^2}\phi_I')'}{\sqrt{\rho^2+m^2}}-\frac{\rho k^2}{(\rho^2+m^2)^{2}}\phi_I - \frac{1}{4\rho}\lambda_{j,m'}\phi_I = 0,
\end{equation}
where the $\lambda_{j,m'}$ are those in Eq.~(\ref{scalarLambda}).  This is the same equation of motion as for the transverse scalar $\delta y$ and so has the same solutions and meson spectrum,
\begin{equation}
-k^2_I=\frac{m^2}{4}(2+j+|m'|+4n)(4+j+|m'|+4n).
\end{equation}
The dual operators have the dimensions $2+\frac{j}{2}$.

Moving on to the Type II fluctuations, we find that the $b=\mu$ equations relate $\phi_{II}$ and $\tilde{\phi}_{II}$ by
\begin{equation}
\label{type2condition}
\frac{(\rho\sqrt{\rho^2+m^2}\phi_{II})'}{\rho\sqrt{\rho^2+m^2}}=\frac{\lambda}{4\rho^2}\tilde{\phi}_{II}.
\end{equation}
Before going on, note that for $j=0$ this equation reduces to $(\rho\sqrt{\rho^2+m^2}\phi_{II})'=0$, which has the solution
\begin{equation}
\phi_{II}\propto\frac{1}{\rho\sqrt{\rho^2+m^2}},
\end{equation}
which is singular at $\rho=0$.  We therefore consider only $j>0$ modes.  Using the condition Eq.~(\ref{type2condition}), the $b=\rho$ and $b=i$ equations are identical,
\begin{equation}
\label{type2eq}
\frac{1}{\rho\sqrt{\rho^2+m^2}}\left[\frac{\rho\sqrt{\rho^2+m^2}}{1-\frac{4m^2\rho^2k^2}{\lambda_{j,m'}(m^2+\rho^2)^2)}}\phi_{II}'\right]'=\frac{\lambda_{j,m'}}{4\rho^2}\phi_{II},
\end{equation}
where the $\lambda_{j,m'}$ are as in Eq.~(\ref{scalarLambda}).  This equation of motion presumably has analytical solutions in terms of hypergeometric functions and an interesting $k$-dependent prefactor.  Unfortunately, we have not been able to find it.  The best we have been able to do is to solve Eq.~(\ref{type2eq}) at $k=0$,
\begin{equation}
\tilde{\phi}_{II,\pm}=\rho^{\pm \frac{m'}{2}}\, _2F_1\left(\frac{-j\pm m'}{4},\frac{2+j\pm m'}{4};1\pm \frac{m'}{2};-\frac{\rho^2}{m^2}  \right).
\end{equation}
At large $\rho$ these solutions go as
\begin{equation}
\tilde{\phi}_{II}\sim \rho^{\frac{j}{2}}+\rho^{-1-\frac{j}{2}},
\end{equation}
so that the dimensions of the dual operators are indeed $2+\frac{j}{2}$.  Nonetheless, the meson spectrum could be obtained by numerically solving Eq.~(\ref{type2eq}); we have done this for various $j,m'$ and found a meson spectrum that agrees with the masses Eq.~(\ref{yMesons}) to approximately one percent.

The last set of modes, the Type III fluctuations, is by far the hardest to solve.  To see this, recall that there are two types of vector harmonics on a round $\mathbb{S}^3$ and therefore on a round $\mathbb{RP}^3\approx\mathbb{S}^3/\mathbb{Z}_2$, those that fall into $\left(\frac{j\pm 1}{2},\frac{j\mp 1}{2}\right)$ representations and the gradients of the scalar harmonics, $D_i\mathcal{Y}^j$, which fall into $\left(\frac{j}{2},\frac{j}{2}\right)$ representations.  There are then three representations that can mix for general $j,m'$: $\left(\frac{j}{2};\frac{j}{2},\frac{m'}{2}\right)$, $\left(\frac{j}{2};\frac{j-2}{2},\frac{m'}{2}\right)$, and $\left(\frac{j}{2};\frac{j+2}{2},\frac{m'}{2}\right)$.  Since
\begin{equation}
D_iD^i(D_k\mathcal{Y}^j)=0
\end{equation}
there are only two linearly independent eigenmode representations for each $j,m'$.  These modes will generally mix rather nastily.  There are some exceptions.  First, for $j=0$ there are only the $\left(0;1,\frac{m'}{2}\right)$ vectors, so these do not mix and are simple to analyze.  Second, for the maximal $|m'|=j+2$ there is only the $\left(\frac{j}{2};\frac{j+2}{2},\frac{m'+2}{2}\right)$ vector.  Finally, the near-maximal case $|m'|=j$ yields a single mode which is a combination of the $\left(\frac{j}{2};\frac{j}{2},\frac{j}{2}\right)$ and $\left(\frac{j}{2};\frac{j+2}{2},\frac{j}{2}\right)$ representations.

It is easy to perform the harmonic analysis for the $j=0$ fields on the internal space.  Unfortunately, we cannot solve the corresponding sectors of meson spectra.  We look at these fields anyway in order to verify a claim in Sec.~\ref{abjmMesons} about their near-boundary expansion.  The eigenvalues are
\begin{equation}
m'=\pm 2: \,\, \lambda=4, \,\,\,\, m'=0:\,\, \lambda=\frac{4}{\left(1+\frac{y^2}{\rho^2}\right)^2}
\end{equation}
and the fields also obey
\begin{equation}
\epsilon_i\,^{kl}\partial_k\mathcal{Y}_{l}^{0}=-2\mathcal{Y}_i^{0}\sqrt{g_{\mathcal{C}_3}}\left\{ \begin{array}{cl}1, & m'=\pm 2, \\ \frac{2}{1+\frac{y^2}{\rho^2}}, & m'=0. \end{array}\right.
\end{equation}
The equations of motion for these fields Eq.~(\ref{eomA}) are then
\begin{eqnarray}
\nonumber
\frac{\left(\rho(\rho^2+m^2)^{3/2}\phi_{III}' \right)'}{\rho(\rho^2+m^2)^{3/2}}+\frac{(2\rho^2-m^2)+\frac{k^2\rho^2}{\rho^2+m^2}}{\rho^2(\rho^2+m^2)}\phi_{III}=0
\end{eqnarray}
for $m'=\pm 2$ and
\begin{equation}
\frac{\rho}{\sqrt{\rho^2+m^2}}\left(\frac{(\rho^2+m^2)^{5/2}}{\rho}\phi_{III}\right)'+(2\rho^2+k^2)\phi_{III}=0
\end{equation}
for $m'=0$.  In each case the large-$\rho$ solution for $\phi_{III}$ is given by
\begin{equation}
\phi_{III}=\frac{a_1}{\rho}+\frac{a_2}{\rho^2}+...
\end{equation}
As discussed in Sec.~\ref{abjmMesons}, the dimension of the dual operator in the supersymmetric theory is $1$ so we should treat $a_2$ as a source.  We do so in Sec.~\ref{abjmRG} and find the one-point function of the dual operator.

\bibliography{refs}

\end{document}